\documentclass[12pt]{article}
\usepackage[utf8]{inputenc}
\usepackage{setspace}
\AtBeginEnvironment{table}{\singlespacing}
\usepackage[margin=1.0in]{geometry}
\usepackage{graphicx}
\usepackage{natbib}
\usepackage{amsmath}
\usepackage{amssymb}
\usepackage{amsfonts}
\usepackage{dsfont}
\usepackage{lineno}
\usepackage{xcolor}
\usepackage{hyperref}
\usepackage{subcaption}
\usepackage{algpseudocode}
\usepackage{algorithm}

\newcommand{\bI}{ {\boldsymbol I} }

\newcommand{\bR}{ {\boldsymbol R} }

\newcommand{\bv}{ {\boldsymbol v} }

\newcommand{\bw}{ {\boldsymbol w} }

\newcommand{\bx}{ {\boldsymbol x} }

\newcommand{\balpha}{ {\boldsymbol \alpha} }
\newcommand{\bbeta}{ {\boldsymbol \beta} }

\newcommand{\bzero}{ {\boldsymbol 0} }

\newcommand{\N}{\mathcal{N}}

\DeclareSymbolFont{bbold}{U}{bbold}{m}{n}
\DeclareSymbolFontAlphabet{\mathbbold}{bbold}

\usepackage{array}
\usepackage{makecell}

\usepackage{letltxmacro}
\LetLtxMacro{\originaleqref}{\eqref}
\renewcommand{\eqref}{Eq.~\originaleqref}

\usepackage{titling}
\predate{}
\postdate{}
\usepackage{authblk}
\title{Hierarchical models for small area estimation using zero-inflated forest inventory variables: comparison and implementation}
\date{} 
\author[1,2,3,*]{Grayson W. White} 
\author[1,2]{Andrew O. Finley} 
\author[4]{Josh K. Yamamoto}
\author[2]{Jennifer L. Green}
\author[5]{Tracey S. Frescino}
\author[1]{David W. MacFarlane}
\author[6]{Hans-Erik Andersen}
\author[7]{Grant M. Domke}
\affil[1]{{\small Department of Forestry, Michigan State University, East Lansing, MI, USA}}
\affil[2]{{\small Department of Statistics \& Probability, Michigan State University, East Lansing, MI, USA}}
\affil[3]{{\small Department of Mathematics \& Statistics, Reed College, Portland, OR, USA}}
\affil[4]{{\small Redcastle Resources, Inc., Salt Lake City, UT, USA}}
\affil[5]{{\small Rocky Mountain Research Station, USDA Forest Service, Riverdale, UT, USA}}
\affil[6]{{\small Pacific Northwest Research Station, USDA Forest Service, Seattle, WA, USA}}
\affil[7]{{\small Northern Research Station, United States Department of Agriculture Forest Service, St. Paul, MN}}
\affil[*]{{\small Corresponding author: Grayson W. White, gwhite@reed.edu}}

\begin{document}
\maketitle

\newpage

\newpage

\begin{abstract} 
National Forest Inventory (NFI) data are typically limited to sparse networks of sample locations due to cost constraints. While design-based estimators provide reliable forest parameter estimates for large areas, there is increasing interest in model-based small area estimation (SAE) methods to improve precision for smaller spatial, temporal, or biophysical domains. SAE methods can be broadly categorized into area- and unit-level models, with unit-level models offering greater flexibility, making them the focus of this study. Ensuring valid inference requires satisfying model distributional assumptions, which is particularly challenging for NFI variables that exhibit positive support and zero-inflation, such as forest biomass, carbon, and volume. Here, we evaluate nine candidate estimators, including two-stage unit-level hierarchical Bayesian models, single-stage Bayesian models, and two-stage frequentist models, for estimating forest biomass at the county level in Nevada and Washington, United States. Estimator performance is assessed using repeated sampling from simulated populations and unit-level cross-validation with FIA data. Results show that small area estimators incorporating a two-stage approach to account for zero-inflation, county-specific random intercepts and residual variances, and spatial random effects yield the most accurate and well-calibrated county-level estimates, with spatial effects providing the greatest benefits when spatial autocorrelation is present in the underlying population.
\end{abstract}


\newpage

\section{Introduction}\label{sec:intro}

National Forest Inventories (NFIs) play a critical role in collecting data and monitoring forest trends to assess resource availability, health, composition, and other economic and ecological attributes across spatial and temporal scales. Traditionally, NFIs have been designed to provide precise estimates at broad spatial scales, such as state- or nation-level assessments of forest attributes like timber volume and biomass. However, there is growing interest in obtaining more precise biomass estimates at finer spatial scales, such as the county level \citep{prisley2021needs, wiener2021united, congress2022}. This rising demand, coupled with the widespread availability of high-resolution remotely sensed data and other auxiliary information, has prompted users to develop and apply innovative small area estimation (SAE) methods that integrate NFI data with remote sensing products \citep{cao2022increased, may2023spatially, finley2024models, Nothdurft2025}.

Despite the wide variety of SAE methods, they can generally be categorized into two main approaches: area-level and unit-level methods \citep{rao15}. Both aim to estimate the same population parameter of interest but differ in their use of data. In area-level estimation, survey unit response variables are aggregated within each area. These aggregates are referred to as direct estimates and are typically generated using a design-based estimator. Direct estimates are then set as the area-level response variable in a regression model that might include area-level summaries of predictor variables and structured random effects. The goal of areal models is to use auxiliary information to smooth noisy direct estimates. In forest inventory, area-level models have been used extensively to estimate forest attributes \citep{cao2022increased, may2023spatially, shannon2024toward}. 

In contrast, unit-level approaches retain response variables at the individual unit (in our case, forest inventory plot) level, making it possible to leverage precise unit locations to estimate fine-scale spatial relationships more effectively. These unit-level attributes are set as the response variable and coupled with spatial and/or temporally aligned predictor variables, possibly along with structured random effects, in a predictive model. This model is then used to predict for all unobserved units, and predictions are subsequently aggregated to any user-defined area of interest. Unit-level models have been applied to forest inventory contexts in a variety of studies \citep{breidenbach2012small, finley2024models, kangas, SHANNON2025122999}. Given their advantages and flexibility, we focus entirely on unit-level models in this study. 

Model-based estimators rely entirely on the assumed data-generating process and the selection of an appropriate model. As a result, particular care must be taken in specifying SAE models and conducting rigorous model checking. One of the most effective approaches to evaluating SAE models is through the use of simulated populations that closely mimic the true---but only partially observed---population of interest. Simulated populations allow for examination of how inference varies under different estimators and for comparisons of resulting estimates against ``true'' values. To ensure that this evaluation is meaningful, it is essential to generate simulated populations using processes that are not closely aligned with the models under assessment \cite{tzavidis2018start, white2024assessing}.

Beyond assessment using simulated populations, models can be evaluated through cross-validation using observed data (e.g., leave-one-out or $k$-fold cross-validation). However, in SAE studies, the primary population parameters of interest exist at the area level. Unit-level models can be assessed using cross-validation at the unit level (i.e., iteratively holdout one or more observations, predict for those holdouts, and compare the predictions to the holdout true values); however, we must be careful to verify that the unit-level assessments align with how well the estimator performs once predictions are aggregated to the desired areas of interest. 

This study evaluates a range of unit-level SAE models for estimating average biomass at the county level in Nevada and Washington, United States (US). A key challenge in this context arises when estimating biomass across areas with a mix of forest and non-forest landcover. Specifically, biomass estimates exhibit a mixture of continuous positive values and true zeros, a phenomenon referred to here as zero-inflation. While the term zero-inflation is commonly used in the statistical literature to describe a discrete distribution with an excessive number of zeros, in this case, biomass follows a continuous distribution with an additional zero component. 

Various model-based approaches have been developed to address zero-inflation in SAE. Notably, \cite{pfeffermann} introduced a two-stage mixture model to account for zero-inflation in the response variable, exploring both frequentist and Bayesian inference. Their findings suggest that mean squared error (MSE) estimation is more straightforward in the Bayesian paradigm because Markov chain Monte Carlo (MCMC) naturally propagates uncertainty across model stages. Expanding on this work, \cite{chandra_sud} applied the same two-stage model in a frequentist setting and introduced a parametric bootstrap-based MSE estimator.

In forest inventory applications, zero-inflation has received relatively limited attention, despite true zeros commonly occurring in forest inventory attributes. \cite{finley2011hierarchical} developed a two-stage model for zero-inflation in continuous forest attributes such as biomass, volume, and age, employing a hierarchical Bayesian framework with Gaussian process-based spatial random effects. Their approach enables unit-level predictions of forest attributes along with uncertainty quantification, though they did not directly produce small area estimates. More recently, \cite{white2024small} applied the zero-inflated SAE model from \cite{chandra_sud} to NFI data in Nevada, generating county-level biomass estimates. Their study compared the zero-inflated estimator to other commonly used small area estimators, including estimators based on the Battese--Harter--Fuller unit-level model and the Fay--Herriot area-level model \citep{fay1979estimates, battese1988error}. Their simulation results indicate the zero-inflated estimator improves point estimates and produces competitive MSE estimates, though further refinements remain possible \citep[see Figure 2 in][]{white2024small}.

In this study, we compare and extend model-based SAE approaches that account for zero-inflation, applying them to data from the US Department of Agriculture (USDA) Forest Service Forest Inventory and Analysis (FIA) Program (the NFI program of the US) paired with remotely sensed auxiliary data products for Nevada and Washington, as described in Section~\ref{sec:data}. Specifically, we evaluate nine model-based approaches, including the zero-inflated estimator from \cite{chandra_sud} and eight hierarchical Bayesian estimators of increasing complexity. These Bayesian estimators include both single-stage models that do not explicitly address zero-inflation and two-stage models designed to account for a preponderance of zeros. With counties within each state defining our small areas of interest, we investigate the effects of incorporating county-varying intercepts, county-varying coefficients, county-specific residual variances, and space-varying intercepts. Section~\ref{sec:estimators} introduces these models and Sections~\ref{sec:freq_two_stage} and~\ref{sec:bayes_two_stage} provide further details on the the frequentist and Bayesian models, respectively. 

Rather than defaulting to the most complex specification, we sequentially introduce model components and evaluate their effect on estimator performance. This approach allows us to identify when added complexity improves estimation and when simpler models suffice. Relative to existing literature, \cite{finley2011hierarchical} considered spatial effects but did not include county-specific terms, while \cite{pfeffermann} and \cite{chandra_sud} did not incorporate spatial dependencies.

In Section~\ref{sec:model_impl}, we introduce nine estimators based on the models introduced previously and discuss details of model implementation. To evaluate the introduced estimators, we conduct a simulation study following the methodology of \cite{white2024assessing} and a FIA data application, as described in Sections~\ref{sec:simulation} and~\ref{sec:fia-data-setup}, respectively. Metrics for comparison of estimators in the simulation study and FIA data application are introduced in Section~\ref{sec:metrics}. Section~\ref{sec:results} presents the simulation results and applies the estimators to FIA data, using cross-validation to assess unit-level model performance. Finally, Section~\ref{sec:discussion} summarizes our findings and discusses their implications, and Section~\ref{sec:conclusion} presents key takeaways and summarizes directions for future research in SAE methods for forest inventory.

\section{Methods}\label{sec:methods}

\subsection{Data}\label{sec:data}

The motivating data are from the FIA Program and consist of inventory plot estimates of live aboveground tree biomass density (Mg/ha). These estimates were obtained from the most recent measurement for each plot in the FIA database, downloaded on February 8, 2023, for the states of Nevada and Washington \citep{burrillforest}. Nevada was selected because it has a distinctive biomass distribution: much of the state's arid environment has little to no biomass, punctuated by sky islands with non-zero forest biomass. Washington was selected because it exhibits large differences in biomass across counties, ranging from very high biomass densities on the Olympic Peninsula to near-zero biomass in counties east of the Cascade Range. At the time of our data query, we accessed the most recent measurement of each FIA plot, with most measurements occurring in the 10-year interval from 2010 to 2019. These data were collected from a panel of plots measured annually across a systematic grid of hexagons approximately 2,500 hectares in size. Biomass estimates were restricted to live trees only and include all trees with diameters of 2.54 cm or greater. This resulted in a sample size of $n$ equal to 11,848 and 7,094 FIA plots for Nevada and Washington, respectively.

Estimators and simulations, described in Sections~\ref{sec:estimators} and \ref{sec:simulation}, respectively, were informed using five auxiliary variables: National Land Cover Dataset Analytical Tree Canopy Cover 2016 (hereafter \texttt{tcc}) \citep{yang2018new}; LANDFIRE 2010 Digital Elevation Model (hereafter \texttt{elev}) \citep{usgs2019ned}; US Geological Survey Terrain Ruggedness Index (hereafter \texttt{tri}) \citep{usgs2019ned}; PRISM mean annual precipitation, 30yr normals (1991–2020) (hereafter \texttt{ppt}) \citep{daly2002knowledge}; and LANDFIRE 2014 tree/non-tree lifeform mask (hereafter \texttt{tnt}) \citep{rollins2009landfire, picotte2019landfire}. The \texttt{tcc} variable is a measure of average tree canopy cover in a given pixel, \texttt{elev} gives elevation, \texttt{tri} gives terrain ruggedness, \texttt{ppt} gives 30-year mean precipitation, and \texttt{tnt} is a binary indicator distinguishing between pixels with and without trees. These auxiliary variables were resampled to 90 m resolution and available wall-to-wall in both states. At locations with FIA plots, these variables were matched with the corresponding plot and then used as predictors in the models' regression components and to inform simulated population generation.

\subsection{Model-based estimation}\label{sec:estimators}

We evaluate nine candidate model-based estimators for estimating average biomass density at the county level across Nevada and Washington. The first estimator employs a frequentist approach using a two-stage regression. The remaining eight adopt a Bayesian framework, incorporating both one-stage and two-stage regression structures. A summary of these estimators is provided in Table~\ref{tab:estimator_desc}.

\begin{table}[ht!]
    \centering
    \begin{tabular}{p{0.35\linewidth} | p{0.6\linewidth}}
      Estimator  & Description \\ \hline
F ZI CVI & A frequentist two-stage estimator. The first stage model is a generalized linear mixed model with a county-varying intercept, and the second stage model is a linear mixed model with a county-varying intercept. \\ 
 B CVI & A Bayesian single-stage estimator based on a linear mixed model with a county-varying intercept.   \\ 
 B CVC & The same as B CVI, but with county-varying coefficients. \\ 
 B ZI CVI & A Bayesian two-stage estimator. The first stage model is a generalized linear mixed model with a county-varying intercept, and the second stage model is a linear mixed model with a county-varying intercept. \\ 
 B ZI CVC & The same as B ZI CVI, but with county-varying coefficients.  \\ 
 B ZI CVI CRV & The same as B ZI CVI, but with county-specific residual variances. \\ 
 B ZI CVC CRV & The same as B ZI CVC, but with county-specific residual variances. \\ 
 B ZI CVI SVI CRV & The same as B ZI CVI CRV, but with an added spatial random effect, modeled as a Nearest Neighbor Gaussian process (NNGP) on the intercept. \\ 
 B ZI CVC SVI CRV & The same as B ZI CVC CRV, but with an added spatial random effect, modeled as a NNGP on the intercept. 
 \end{tabular}
 \caption{\label{tab:estimator_desc} Description of the candidate models considered for estimating county-level forest biomass. Abbreviations are: frequentist (F); Bayesian (B); zero-inflated (ZI); county-varying intercept (CVI); county-varying coefficient (CVC); county-specific residual variance (CRV); space-varying intercept (SVI).}
\end{table}

To improve adherence to normality assumptions and ensure positive support for back-transformed predictions, models were fit using transformed response variables. While logarithmic transformation of non-zero biomass is common in such contexts, we found it overly aggressive in this application, producing left-skewed transformed distributions. Instead, we applied root-based transformations, selecting state-specific powers to better align with regional biomass distributions: a fourth-root transformation for Washington and a square-root transformation for Nevada. In exploring transformation options, we focused on the distribution of non-zero response values, assuming that the zero-inflated model components would account for excess zeros. Compared to log-based approaches, root-based transformations offer the added benefit that zeros remain unchanged on both the transformed and back-transformed scales.

Although all models are fit at the unit level, our primary inferential target is the average biomass density at the county level, denoted $\mu_j$, where $j$ indexes counties within each state. Let $\ell$ denote the spatial location of a plot in county $j$ (i.e., $\ell \in j$). Two response variables are modeled: $z(\ell)$, a Bernoulli indicator of biomass presence ($z(\ell) = 1$ for non-zero biomass, $z(\ell) = 0$ otherwise), and $y(\ell)$, the (transformed) continuous biomass value at plots where $z(\ell) = 1$. Predictors are indexed by location as $\bx(\ell)$ and $\bv(\ell)$, with random effects defined at the county level. This notation also facilitates the introduction of a continuous spatial process random effect, $w(\ell)$, in later models. The relationship between $\ell$ and $j$ is assumed throughout and is not restated in each model. In the sections that follow, we describe the candidate unit-level models in detail and outline how each is used to estimate these small area population parameters of interest.

\subsection{Models for frequentist two-stage estimation}\label{sec:freq_two_stage}

The frequentist estimator (F ZI CVI) adopts a two-stage framework originally developed by \cite{chandra_sud} and later applied to forest inventory data by \cite{white2024small}. The first stage models biomass presence using a Bernoulli mixed model with county-level random intercepts, while the second stage models transformed non-zero biomass values with a linear mixed model, also including county-level random intercepts. Together, these stages provide an estimator and a corresponding MSE estimator suitable for zero-inflated continuous responses. 

The continuous response model is
\begin{equation}
    y(\ell) = \beta_0 + \tilde\beta_{0,j} + \bx(\ell)^\top\bbeta + \varepsilon(\ell),
    \label{mod:freq_two_state_y}
\end{equation}
where $\beta_0$ is the intercept, $\tilde\beta_{0,j}\overset{\text{iid}}{\sim} \N(0, \sigma^2_{\tilde\beta_0})$, $\bx(\ell)$ is a $p\times1$ vector of predictors with regression coefficients $\bbeta$, and $\varepsilon(\ell)\overset{\text{iid}}{\sim} \N(0, \tau^2)$. 

The Bernoulli stage for biomass presence is
\begin{equation}
    \log\left(\frac{p(\ell)}{1 - p(\ell)}\right) = \alpha_0 + \tilde\alpha_{0,j} + \bv(\ell)^\top\balpha,
    \label{mod:freq_two_state_p}
\end{equation}
where $p(\ell)$ denotes the probability of non-zero biomass, i.e., $p(\ell) = \Pr(z(\ell)=1)$, where $z(\ell)$ is the Bernoulli indicator defined above, $\alpha_0$ is the intercept, $\tilde\alpha_{0,j}\overset{\text{iid}}{\sim} \N(0, \sigma^2_{\tilde\alpha_0})$, and $\balpha$ is a $q\times1$ coefficient vector for predictors $\bv(\ell)$. 

F ZI CVI provides a useful benchmark for handling zero-inflated continuous responses with county-level effects. However, uncertainty quantification and extension to more complex model structures can be cumbersome in the frequentist setting. To address these challenges, we next consider a sequence of Bayesian estimators that build from a common unit-level regression framework. These models begin with a simple county-varying intercept specification and progressively incorporate county-varying slopes, explicit two-stage formulations for zero-inflation, county-specific residual variances, and spatial random effects.

\subsection{Models for Bayesian estimation}\label{sec:bayes_two_stage}

The Bayesian estimators summarized in Table~\ref{tab:estimator_desc} extend the frequentist framework by incorporating hierarchical modeling and MCMC-based inference. All models share a common unit-level regression structure but vary in how they address zero-inflation, county-level heterogeneity, and spatial dependence. To clarify the role of each modeling feature, we present the estimators in order of increasing complexity, beginning with the baseline B CVI model and sequentially adding components.

All parameters are estimated unless otherwise noted. Prior distributions and hyperparameter values are given in Table~\ref{tab:prior_distributions}. Priors were chosen to be weakly informative and commonly used in hierarchical modeling, with the goal of avoiding undue influence on posterior inference.

\paragraph{B CVI.} 
The simplest model is the county-varying intercept specification, the Bayesian analogue to \eqref{mod:freq_two_state_y},
\begin{equation}
    y(\ell) = \beta_0 + \tilde\beta_{0,j} + \bx(\ell)^\top\bbeta + \varepsilon(\ell),
    \label{mod:bayes_CVI}
\end{equation}
with $\beta_0 \sim \N(0, \sigma^2_{\beta_0})$, 
$\tilde\beta_{0,j} \overset{\text{iid}}{\sim} \N(0, \sigma^2_{\tilde\beta_0})$, 
$\bbeta \sim \N(\bzero, \sigma^2_{\beta}\bI_p)$, and 
$\varepsilon(\ell)\overset{\text{iid}}{\sim}\N(0,\tau^2)$.  
This model allows county-level differences in mean biomass while assuming predictor effects are constant across counties.

\paragraph{B CVC.} 
The B CVC model extends B CVI to allow regression coefficients to vary across counties, so that
\begin{equation}
    y(\ell) = \beta_0 + \tilde\beta_{0,j} + \bx(\ell)^\top\big(\bbeta + \tilde\bbeta_j\big) + \varepsilon(\ell),
    \label{mod:bayes_CVC}
\end{equation}
where $\tilde\bbeta_j = (\tilde\beta_{1,j},\ldots,\tilde\beta_{p,j})^\top$ with 
$\tilde\beta_{k,j} \overset{\text{iid}}{\sim}\N(0, \sigma^2_{\tilde\beta_k})$ for $k=1,\ldots,p$.  
This flexibility captures differences in how predictors relate to biomass across counties.

Moving to zero-inflated models, the Bayesian analogue of the frequentist Bernoulli stage in \eqref{mod:freq_two_state_p}, used for all two-stage Bayesian models, is
\begin{equation}
    \log\left( \frac{p(\ell)}{1 - p(\ell)} \right) = \alpha_0 + \tilde\alpha_{0,j} + \bv(\ell)^\top\balpha,
    \label{mod:bayes_bern}
\end{equation}
with $\alpha_0 \sim \N(0, \sigma^2_{\alpha_0})$, 
$\tilde\alpha_{0,j} \overset{\text{iid}}{\sim} \N(0, \sigma^2_{\tilde\alpha_0})$, and 
$\balpha \sim \N(\bzero, \sigma^2_{\alpha}\bI_q)$.  
All zero-inflated models use \eqref{mod:bayes_bern} to define $z(\ell)$.

\paragraph{B ZI CVI.} 
The first two-stage Bayesian model mirrors the frequentist zero-inflated specification in Section~\ref{sec:freq_two_stage}. The continuous response is modeled conditional on $z(\ell)$ as
\begin{equation}
    y(\ell) = z(\ell)\left(\beta_0 + \tilde\beta_{0,j} + \bx(\ell)^\top\bbeta + \varepsilon_1(\ell)\right) + (1 - z(\ell))\varepsilon_2,
    \label{mod:bayes_zi_CVI}
\end{equation}
where $\varepsilon_1(\ell)\overset{\text{iid}}{\sim}\N(0,\tau_1^2)$ and $\varepsilon_2\overset{\text{iid}}{\sim}\N(0,\tau_2^2)$, with $\tau_2^2$ fixed at $10^{-6}$ following \cite{finley2011hierarchical}. This model accommodates excess zeros, while retaining a county-varying intercept structure.

\paragraph{B ZI CVC.} 
B ZI CVI is extended to allow predictor effects to vary across counties, giving
\begin{equation}
    y(\ell) = z(\ell)\left(\beta_0 + \tilde\beta_{0,j} + \bx(\ell)^\top\!\big(\bbeta + \tilde\bbeta_j\big) + \varepsilon_1(\ell)\right) + (1 - z(\ell))\varepsilon_2,
    \label{mod:bayes_zi_CVC}
\end{equation}
which is useful when relationships between biomass and predictors differ across counties.

\paragraph{B ZI CVI CRV.} 
Residual variance often increases with mean biomass, producing heteroscedasticity across counties. To address this, B ZI CVI is extended to allow county-specific residual variances so that
\begin{equation}
    y(\ell) = z(\ell)\left(\beta_0 + \tilde\beta_{0,j} + \bx(\ell)^\top\bbeta + \varepsilon_{1,j}(\ell)\right) + (1 - z(\ell))\varepsilon_2,
    \label{mod:bayes_zi_CVI_CRV}
\end{equation}
with $\varepsilon_{1,j}(\ell)\sim\N(0,\tau_{1,j}^2)$.  

\paragraph{B ZI CVC CRV.} 
Combining the previous two extensions yields a model with both county-varying coefficients and county-specific residual variances, expressed as
\begin{equation}
    y(\ell) = z(\ell)\left(\beta_0 + \tilde\beta_{0,j} + \bx(\ell)^\top\!\big(\bbeta + \tilde\bbeta_j\big) + \varepsilon_{1,j}(\ell)\right) + (1 - z(\ell))\varepsilon_2,
    \label{mod:bayes_zi_CVC_CRV}
\end{equation}
which accounts for county-varying relationships between predictors and biomass and heteroscedasticity in the residuals.

\paragraph{B ZI CVI SVI CRV.} 
Residual spatial variation in biomass can arise from disturbance history, climate gradients, or other unobserved spatial drivers. To capture these, a spatial random effect is added to the intercept, leading to
\begin{equation}
    y(\ell) = z(\ell)\left(\beta_0 + \tilde\beta_{0,j} + \bx(\ell)^\top\bbeta + w(\ell) + \varepsilon_{1,j}(\ell)\right) + (1 - z(\ell))\varepsilon_2,
    \label{mod:bayes_zi_CVI_SVI_CRV}
\end{equation}
where $w(\ell)$ is a spatial random effect that adjusts the intercept based on residual spatial dependence.  
We estimate $w(\ell)$ using the Nearest Neighbor Gaussian Process (NNGP; \citealt{datta2016hierarchical, finley2019efficient}), which provides substantial improvements in run time with negligible differences in inference and prediction compared to a full Gaussian process. 
In brief, the vector of random effects $\bw = (w(\ell_1),\ldots,w(\ell_n))^\top$ is distributed multivariate normal with mean zero and covariance $\sigma^2_w\bR(\phi)$, where $\bR(\phi)$ is the NNGP-derived correlation matrix based on an exponential correlation function with decay parameter $\phi$.

\paragraph{B ZI CVC SVI CRV.} 
The most general Bayesian estimator considered combines all features---county-varying coefficients, county-specific residual variances, and a spatial random effect on the intercept---so that
\begin{equation}
    y(\ell) = z(\ell)\left(\beta_0 + \tilde\beta_{0,j} + \bx(\ell)^\top\!\big(\bbeta + \tilde\bbeta_j\big) + w(\ell) + \varepsilon_{1,j}(\ell)\right) + (1 - z(\ell))\varepsilon_2,
    \label{mod:bayes_zi_CVC_SVI_CRV}
\end{equation}
where $w(\ell)$ is defined as in \eqref{mod:bayes_zi_CVI_SVI_CRV}.  
This specification offers the greatest flexibility, simultaneously addressing zero-inflation, county-level heterogeneity, heteroscedasticity, and residual spatial dependence.

\subsection{Model implementation and comparison}\label{sec:model_impl}

\subsubsection{Frequentist two-stage estimator}\label{sec:freq_pred}

The \texttt{R} \citep{R} package \texttt{saeczi} \citep{saeczi} implements methods presented by \cite{chandra_sud} to fit the F ZI CVI model. However, in order to implement back-transformation of the response variable and bias correction due to back-transformation, we implement the model without direct use of \texttt{saeczi}. Instead, model parameters for F ZI CVI described in Section~\ref{sec:freq_two_stage} were estimated using restricted maximum likelihood via \texttt{lme4} \citep{bates2015package}. 

As described in Section~\ref{sec:intro}, to generate estimates for a small area of interest, we predict biomass and probability of non-zero biomass using \eqref{mod:freq_two_state_y} and \eqref{mod:freq_two_state_p}, respectively, over a fine grid of prediction locations. For generic prediction location $\ell^\ast$ in county $j$ these predictions are
\begin{align}
    \hat{y}^\ast(\ell^\ast) &= \hat{\beta}_0 + \hat{\tilde\beta}_{0,j} + \bx(\ell^\ast)^\top \hat\bbeta, \\
    \hat{p}(\ell^\ast) &= \frac{\exp\left( \hat\alpha_0 + \hat{\tilde\alpha}_{0,j} + \bv(\ell^\ast)^\top \hat\balpha \right)}{1 + \exp\left( \hat\alpha_0 + \hat{\tilde\alpha}_{0,j} + \bv(\ell^\ast)^\top \hat\balpha \right)},
\end{align}
where the hats indicate point estimates. The estimate for $\mu_j$ is then the average of the product of these predictions over the grid of prediction locations
\begin{equation}
    \hat \mu_j = \frac{1}{n^\ast_j}\sum_{\ell^\ast\in U_j} g^{-1}\!\left(\hat{y}^\ast(\ell^\ast)\right) \hat{p}(\ell^\ast),
    \label{mod:freq_prediction}
\end{equation}
where $g^{-1}(\cdot)$ is the bias-corrected inverse of the function used to transform the response variable when fitting the model, and $U_j$ is the set of $n^\ast_j$ prediction locations within county $j$. To correct for bias introduced from applying the inverses of the root-based transformations we follow \cite{gregoire2008regression} and apply bias corrections based on the $n$th noncentral moment of the Gaussian distribution for the $n$th root transformation. In our case, the square- and fourth-root bias-corrected back-transformations are 
\begin{equation}
    g^{-1}(\hat{y}^\ast(\ell^\ast)) = \hat{y}^\ast(\ell^\ast)^2 + \hat\tau^2,
\end{equation}
and 
\begin{equation}
    g^{-1}(\hat{y}^\ast(\ell^\ast)) = \hat{y}^\ast(\ell^\ast)^4 + 6 \hat{y}^\ast(\ell^\ast)^2 \hat\tau^2 + 3\hat\tau^4,
\end{equation}
respectively. The MSE of $\hat\mu_j$ is estimated via a parametric bootstrap introduced by \cite{chandra_sud} and explored further in \cite{white2024small}.

\subsubsection{Bayesian estimators}\label{sec:bayes_pred}

Parameter inference for the Bayesian models in Section~\ref{sec:bayes_two_stage} was based on MCMC. Gibbs and Metropolis–Hastings steps were implemented in \texttt{C++} to efficiently sample from posterior distributions. Code, additional information about the algorithms, and example analyses using simulated data are given in \cite{ZICode}. A list of prior distributions and hyperparameter values is provided in Appendix~\ref{sec:appendixA}. Posterior inference is based on $M=3{,}000$ thinned post-convergence samples (1,000 from each of three chains), following convergence diagnostics and thinning rules from \cite{gelman2013bayesian}. 

Inference about biomass at prediction locations and subsequent county-level means is based on samples from the posterior predictive distribution. For example, under the most general B ZI CVC SVI CRV model (\eqref{mod:bayes_zi_CVC_SVI_CRV}), for a generic prediction location $\ell^\ast$ in county $j$ we generate one posterior predictive sample per retained MCMC iteration $s=1,\ldots,M$. Specifically,
\begin{equation}
 z^{(s)}(\ell^\ast) \sim \text{Bernoulli}\!\left(\frac{\exp\left( \alpha_0^{(s)} + \tilde\alpha_{0,j}^{(s)} + \bv(\ell^\ast)^\top \balpha^{(s)} \right)}{1 + \exp\left( \alpha_0^{(s)} + \tilde\alpha_{0,j}^{(s)} + \bv(\ell^\ast)^\top \balpha^{(s)} \right)}\right),
\end{equation}
and, given $z^{(s)}(\ell^\ast)$,
\begin{multline}
 y^{(s)}(\ell^\ast) \sim \N \Bigl(z^{(s)}(\ell^\ast)\bigl(\beta_0^{(s)} + \tilde\beta_{0,j}^{(s)} + \bx(\ell^\ast)^\top\!\big(\bbeta^{(s)} + \tilde\bbeta_j^{(s)}\big) + w^{(s)}(\ell^\ast)\bigr), \\
 z^{(s)}(\ell^\ast)\tau^{2(s)}_{1,j} + \bigl(1 - z^{(s)}(\ell^\ast)\bigr)\tau^{2(s)}_2 \Bigr).
  \label{mod:bayes_y_ppd}
\end{multline}

Given $M$ posterior predictive samples from \eqref{mod:bayes_y_ppd}, county-level averages are computed as
\begin{equation}
\mu^{(s)}_j = \frac{1}{n^\ast_j}\sum_{\ell^\ast\in U_j} h^{-1}\!\left(y^{(s)}(\ell^\ast)\right),
 \label{mod:bayes_mu_ppd}
\end{equation}
for $s=1,\ldots,M$ and where $h^{-1}(\cdot)$ is the inverse of the function used to transform the response variable when fitting the model. Unlike the frequentist approach where we back-transform the estimated mean of the response variable, $y(\ell^\ast)$, in the Bayesian approach we back-transform the posterior predictive samples, $y^{(s)}(\ell^\ast)$, making any bias correction unnecessary \citep{stow2006bayesian}. Posterior means and credible intervals for $\mu_j$ are then obtained directly from these samples, with the Bayesian analogue of \eqref{mod:freq_prediction} given by
\begin{equation}
\hat{\mu}_j = \frac{1}{M}\sum_{s=1}^M \mu^{(s)}_j.
\end{equation}

The MSE of $\hat\mu_j$ is estimated from the variance of the posterior distribution of $\mu_j$. 

\subsubsection{Variable selection}

The five auxiliary variables described in Section~\ref{sec:data} were used throughout the analysis. In particular, they informed the generation of simulated populations, were included as predictors when fitting models to those populations, and were subsequently applied in the analysis of FIA data. 

For simulated population generation, variable selection was informed by domain knowledge about regional biomass drivers. In both Nevada and Washington, \texttt{tcc} and \texttt{elev} were selected as core predictors due to their expected influence on biomass distribution. In Nevada, we additionally included \texttt{tri} to reflect the ecological relevance of sky islands in structuring biomass presence, while in Washington we included \texttt{ppt} to capture the east--west precipitation gradient across the Cascade Range, which strongly influences biomass patterns. To ensure distinct imputation strategies for different vegetation conditions, we stratified the simulated populations in both states by \texttt{tnt}.

For the candidate estimators applied to both simulated and FIA data, we used the same predictor variables within each state to maintain information consistency with the corresponding data-generating processes. Initially, \texttt{tnt} was excluded from the models. While this exclusion had little effect in Nevada, in Washington it led to poor predictive performance in the Bernoulli models. Consequently, we included \texttt{tnt} as a predictor in all Bernoulli models for Washington.

Table~\ref{tab:predictor_use} summarizes the use of auxiliary variables across model types and data sources.

\begin{table}[!ht]
    \centering
    \begin{tabular}{l|c c c}
    \hline
      Predictor  & Gaussian  & Bernoulli  & Simulated population \\ \hline
      \texttt{tcc} & WA, NV & WA, NV & WA, NV \\
      \texttt{elev} & WA, NV & WA, NV & WA, NV \\
      \texttt{tri} & NV & NV & NV \\
      \texttt{ppt} & WA & WA & WA \\
      \texttt{tnt} & none & WA & WA, NV
 \end{tabular}
 \caption{Auxiliary variables used in model fitting and simulated population generation.}\label{tab:predictor_use} 
\end{table}

\subsection{Simulation study setup}\label{sec:simulation}

We conducted a simulation study to evaluate the performance of the estimators introduced in Section~\ref{sec:estimators}. Simulation studies are especially valuable in SAE contexts because they enable comparisons against known population values, providing direct insight into estimator accuracy and uncertainty quantification. To ensure a fair and realistic evaluation, we followed the methodology proposed by \cite{white2024assessing}.

Simulated populations were constructed using a $k$-nearest neighbors ($k$NN) imputation algorithm applied to auxiliary variables, with neighbors weighted by bootstrap inclusion probabilities. This approach imputes forest inventory attributes for every pixel in the population, yielding a synthetic yet realistic population over which true area-level parameters (e.g., county means) are known. The procedure is described in detail in Algorithm 1 of \cite{white2024assessing}.

We generated separate simulated populations for Nevada and Washington. As shown in Table~\ref{tab:predictor_use}, the $k$NN algorithm used \texttt{tcc}, \texttt{elev}, and \texttt{tri} in Nevada, and \texttt{tcc}, \texttt{elev}, and \texttt{ppt} in Washington. In both states, \texttt{tnt} was used to stratify between treed and non-treed areas. All predictors were centered and scaled before matching to ensure balanced influence. Population generation was implemented using the \texttt{kbaabb} R package \citep{kbaabbrpkg}.

To evaluate estimator performance, we drew $d = 100$ simple random samples from each simulated population. Each sample was selected at the state level, with the number of observations per county matched to the number of FIA plots observed in that county. This preserved the original FIA sampling intensity across space. Estimators were applied to each sample, and performance metrics were computed across the $d$ samples.

\subsection{FIA data analysis setup}\label{sec:fia-data-setup}

In addition to the simulation study, we evaluated estimator performance using observed FIA data in Nevada and Washington through a predictive validation approach based on 10-fold cross-validation. This complements the simulation study by evaluating model behavior under realistic conditions, without assuming a known population generating process.

Cross-validation was performed at the unit level. FIA plots were partitioned into 10 folds, and each model was fit to 9 folds and used to predict on the remaining (held-out) fold. This process was repeated until each fold had served as a test set. For each held-out observation, predictions were generated on the original response scale using the model-specific inverse transformations. For frequentist models, we applied the bias-corrected inverse transformation $g^{-1}(\cdot)$ to the fitted values. For Bayesian models, posterior predictive samples were transformed using the inverse transformation $h^{-1}(\cdot)$, and posterior predictive means were used as point predictions. This procedure produced a complete set of unit-level predictions for each estimator, which were then used to evaluate performance across folds.

\subsection{Metrics for comparison}\label{sec:metrics}

Estimator performance was evaluated using two complementary approaches: repeated samples from simulated populations (Section~\ref{sec:simulation}) and 10-fold cross-validation on FIA data (Section~\ref{sec:fia-data-setup}). In both settings, we assessed performance using root mean square error (RMSE), bias, and coverage, applied to either county-level estimates or unit-level predictions depending on the context.

\subsubsection{Simulation study}

In the simulation study, the focus is on county-level parameters. Let $\mu_j$ denote the true county-level mean biomass in county $j$, and let $\hat{\mu}_{ji}$ be the estimate from the $i$th sample, for $i = 1,\ldots,d$, where $d$ is the number of simple random samples drawn from the simulated population.

The root mean square error (RMSE) for county $j$ is
\begin{equation}
\mathrm{RMSE}(\hat{\mu}_j) = \sqrt{\frac{1}{d} \sum_{i=1}^{d} \left(\hat{\mu}_{ji} - \mu_j\right)^2}.
\end{equation}

Bias is defined as the difference between the mean of the estimates and the true value,
\begin{equation}
\mathrm{Bias}(\hat{\mu}_j) = \frac{1}{d}\sum_{i=1}^d \hat{\mu}_{ji} - \mu_j.
\end{equation}

We also evaluated the bias of the RMSE estimator $\widehat{\mathrm{RMSE}}(\hat{\mu}_j)$ relative to the empirical RMSE defined above. For county $j$, this is
\begin{equation}
\mathrm{Bias}\bigl(\widehat{\mathrm{RMSE}}(\hat{\mu}_{j})\bigr) = \frac{1}{d} \sum_{i=1}^d \widehat{\mathrm{RMSE}}(\hat\mu_{ji}) - \mathrm{RMSE}(\hat{\mu}_j),
\end{equation}
where $\widehat{\mathrm{RMSE}}(\hat\mu_{ji})$ is the estimated RMSE for county $j$ from the $i$th sample.

Coverage is the proportion of estimated uncertainty intervals that contain the true value. Let $I_{ji}$ be the uncertainty interval for $\hat{\mu}_{ji}$. Coverage for county $j$ is
\begin{equation}
\mathrm{Coverage}(\hat{\mu}_j) = \frac{1}{d}\sum_{i=1}^{d} \mathbbold{1}_{I_{ji}}(\mu_j),
\end{equation}
where the indicator function is 1 if $\mu_j \in I_{ji}$ and 0 otherwise. For frequentist estimators, $I_{ji}$ is defined as $\hat{\mu}_{ji} \pm 1.96 \cdot \widehat{\mathrm{RMSE}}(\hat{\mu}_{ji})$. For Bayesian estimators, $I_{ji}$ is based on the 0.025 and 0.975 posterior quantiles of $\mu_j$.

\subsubsection{FIA data analysis}

For the FIA data, performance was evaluated at the unit level using 10-fold cross-validation. Let $y(\ell^\ast)$ denote the observed biomass value at held-out location $\ell^\ast$ and $\hat{y}(\ell^\ast)$ the corresponding model prediction.

The root mean square prediction error (RMSPE) for all held-out locations $\ell^\ast$ is
\begin{equation}
\mathrm{RMSPE}(\hat{y}(\ell^\ast)) = \sqrt{\frac{1}{n}\sum_{i=1}^n \left(\hat{y}(\ell^\ast_i) - y(\ell^\ast_i)\right)^2},
\end{equation}
where $n$ is the number of held-out locations (all locations are held out once over the 10 folds), $\hat{y}(\ell^\ast)$ is given by $g^{-1}(\hat{y}^\ast(\ell^\ast))\hat{p}(\ell^\ast)$ for the frequentist estimator and by the posterior predictive mean $\frac{1}{M}\sum_{s=1}^{M} h^{-1}(y^{(s)}(\ell^\ast))$ for the Bayesian estimators.

Bias is computed across all held-out locations $\ell^\ast$ as
\begin{equation}
\mathrm{Bias}(\hat{y}(\ell^\ast)) = \frac{1}{n}\sum_{i=1}^n \left(\hat{y}(\ell^\ast_i) - y(\ell^\ast_i)\right).
\end{equation}

Coverage is based on the proportion of credible intervals that contain the true held-out value. For Bayesian estimators, the interval is
\begin{equation*}
I_i = \left[ Q_{0.025}\big(h^{-1}(y(\ell^\ast_i))\big),~ Q_{0.975}\big(h^{-1}(y(\ell^\ast_i))\big) \right].
\end{equation*}

Frequentist unit-level prediction intervals were not computed, as the bootstrap procedure from \cite{chandra_sud} is not designed for this purpose.

\section{Results}\label{sec:results}


\subsection{Simulation Study}\label{sec:simulation-res}

The nine estimators introduced in Section~\ref{sec:methods} were evaluated using repeated simple random samples drawn from the simulated populations described in Section~\ref{sec:simulation}. Model fitting was conducted separately for Nevada and Washington to reflect state-specific differences in data-generating processes. Estimator performance varied between states, likely due to ecological contrasts: Nevada’s forests are confined to high-elevation sky islands, while Washington’s forests are concentrated west of the Cascade Range, where precipitation and productivity are higher.

\begin{figure}[ht!]
    \includegraphics[width=\textwidth]{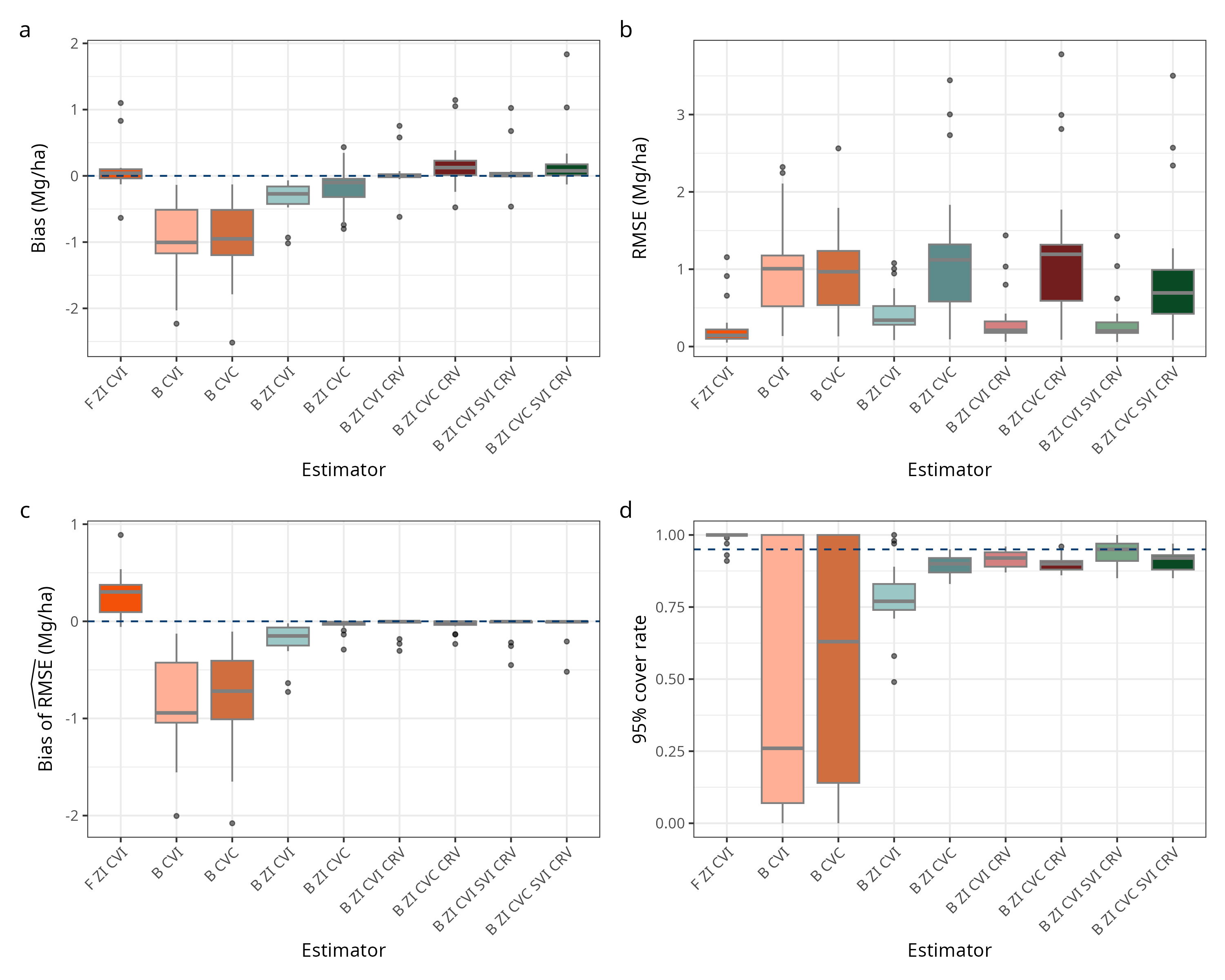}
    \caption{Estimator performance metrics in Nevada. The x-axis and fill correspond to estimator, and the y-axis corresponds to the value of the performance metric. Each point represents the performance metric in a given county for a particular estimator. (a) empirical bias, (b) empirical root mean square error (RMSE), (c) bias of the RMSE estimator, and (d) empirical 95\% uncertainty interval coverage. Abbreviations: root mean square error (RMSE); frequentist (F); zero-inflated (ZI); county-varying intercept (CVI); Bayesian (B); county-varying coefficient (CVC); county-specific residual variance (CRV); space-varying intercept (SVI).}
    \label{fig:nevada}
\end{figure}

For Nevada, Figure~\ref{fig:nevada}a shows that single-stage models have the largest empirical bias. Estimators that include county-specific residual variances perform better than those that omit them, and incorporating county-varying coefficients typically increases bias. The least biased estimators are B ZI CVI CRV and B ZI CVI SVI CRV, which yield nearly identical results.

In Figure~\ref{fig:nevada}b, single-stage estimators B CVI and B CVC, as well as two-stage estimators that include CVC terms, consistently show the largest RMSE. The two-stage estimators with county-varying intercepts and county-specific residual variance terms, along with the F ZI CVI, produce the smallest RMSE, followed closely by B ZI CVI. 

Figure~\ref{fig:nevada}c shows that B CVI and B CVC underestimate RMSE, consistent with poor model specification. B ZI CVI also underestimates RMSE despite accounting for zero inflation, while F ZI CVI consistently overestimates it. The B ZI models with CRV and/or SVI terms are generally accurate but occasionally produce negative outliers.

Figure~\ref{fig:nevada}d illustrates uncertainty interval coverage. B CVI and B CVC show widespread undercoverage (a result of their large negative RMSE bias), while F ZI CVI consistently displays overcoverage (a function of its large positive RMSE bias). Estimators incorporating zero inflation, CRV, and SVI perform best overall, with B ZI CVI SVI CRV achieving median coverage closest to 0.95.

\begin{figure}[!ht]
    \includegraphics[width=\textwidth]{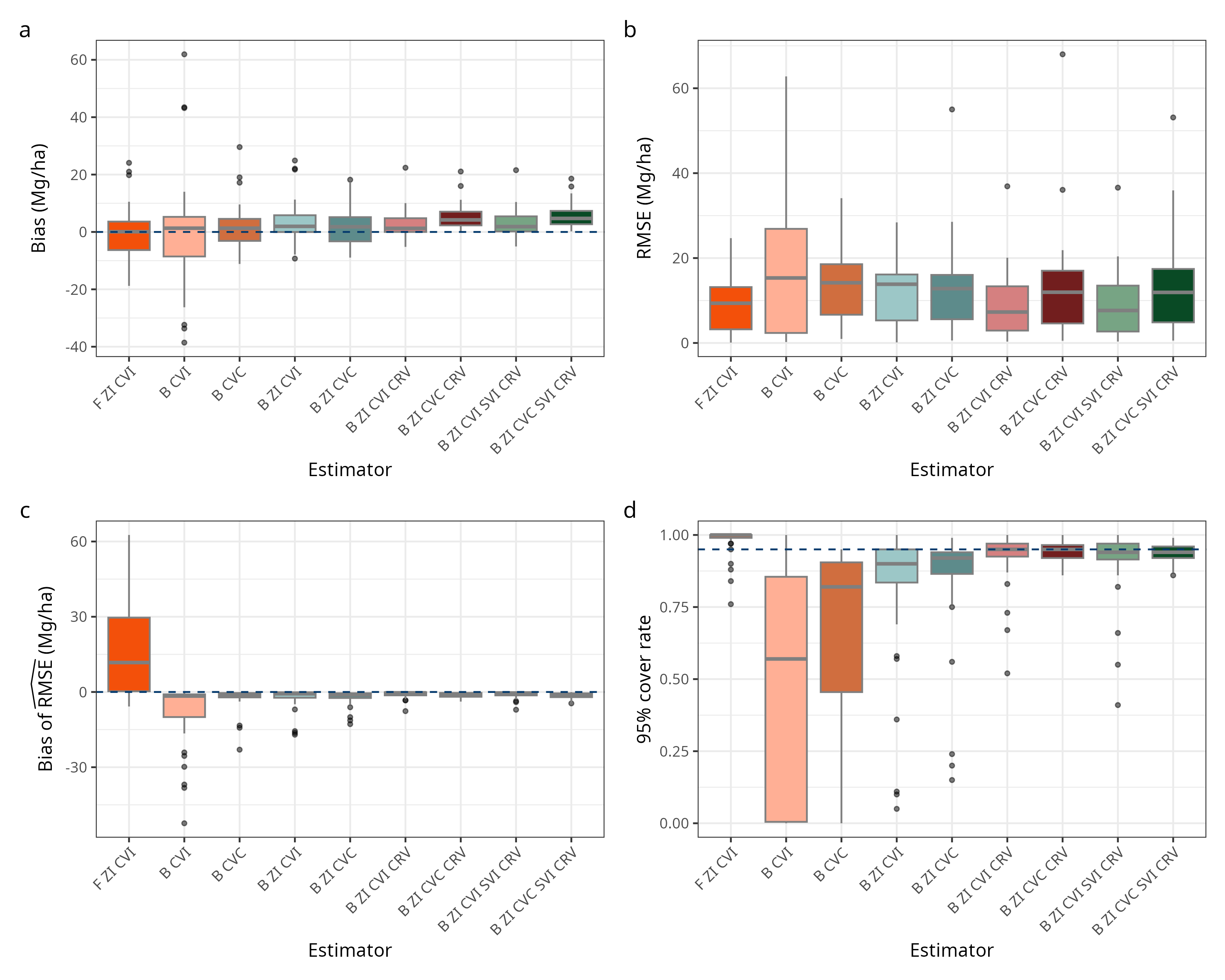}
    \caption{Estimator performance metrics in Washington. The x-axis and fill correspond to estimator, and the y-axis corresponds to the value of the performance metric. Each point represents the performance metric in a given county for a particular estimator. (a) empirical bias, (b) empirical root mean square error (RMSE), (c) bias of the RMSE estimator, and (d) empirical 95\% uncertainty interval coverage. Abbreviations: root mean square error (RMSE); frequentist (F); zero-inflated (ZI); county-varying intercept (CVI); Bayesian (B); county-varying coefficient (CVC); county-specific residual variance (CRV); space-varying intercept (SVI).}
    \label{fig:washington}
\end{figure}

For Washington, Figure~\ref{fig:washington}a shows that the estimators incorporating the two-stage structure all perform similarly. Here too, even the single-stage estimators perform well; however, B CVI shows a marginally larger range of values compared with other models.

In Figure~\ref{fig:washington}b, the F ZI CVI, B ZI CVI CRV, and B ZI CVI SVI CRV estimators have the lowest median RMSE values. Other two-stage Bayesian estimators perform similarly, with single-stage estimators performing slightly worse, particularly B CVI.

Figure~\ref{fig:washington}c shows that F ZI CVI and B CVI perform poorly when estimating their own variability, whereas most other estimators---particularly those with CRV terms---have minimal bias and accurately estimate RMSE.

Figure~\ref{fig:washington}d shows that B CVI and B CVC have the lowest coverage rates, while all two-stage models provide better coverage. Those incorporating both CVC effects and CRV terms, such as B ZI CVC SVI CRV, exhibit the best performance. F ZI CVI shows substantial overcoverage, again a function of its large positive RMSE bias.

\subsection{FIA data application}\label{sec:data-app}

Estimator performance for FIA data was evaluated using 10-fold cross-validation. Based on these results and the simulation study, the B ZI CVI SVI CRV model---which performed well in both analyses---was then used to generate county-level biomass estimates.

\begin{figure}[!ht]
\centering
    \includegraphics[width=0.75\textwidth]{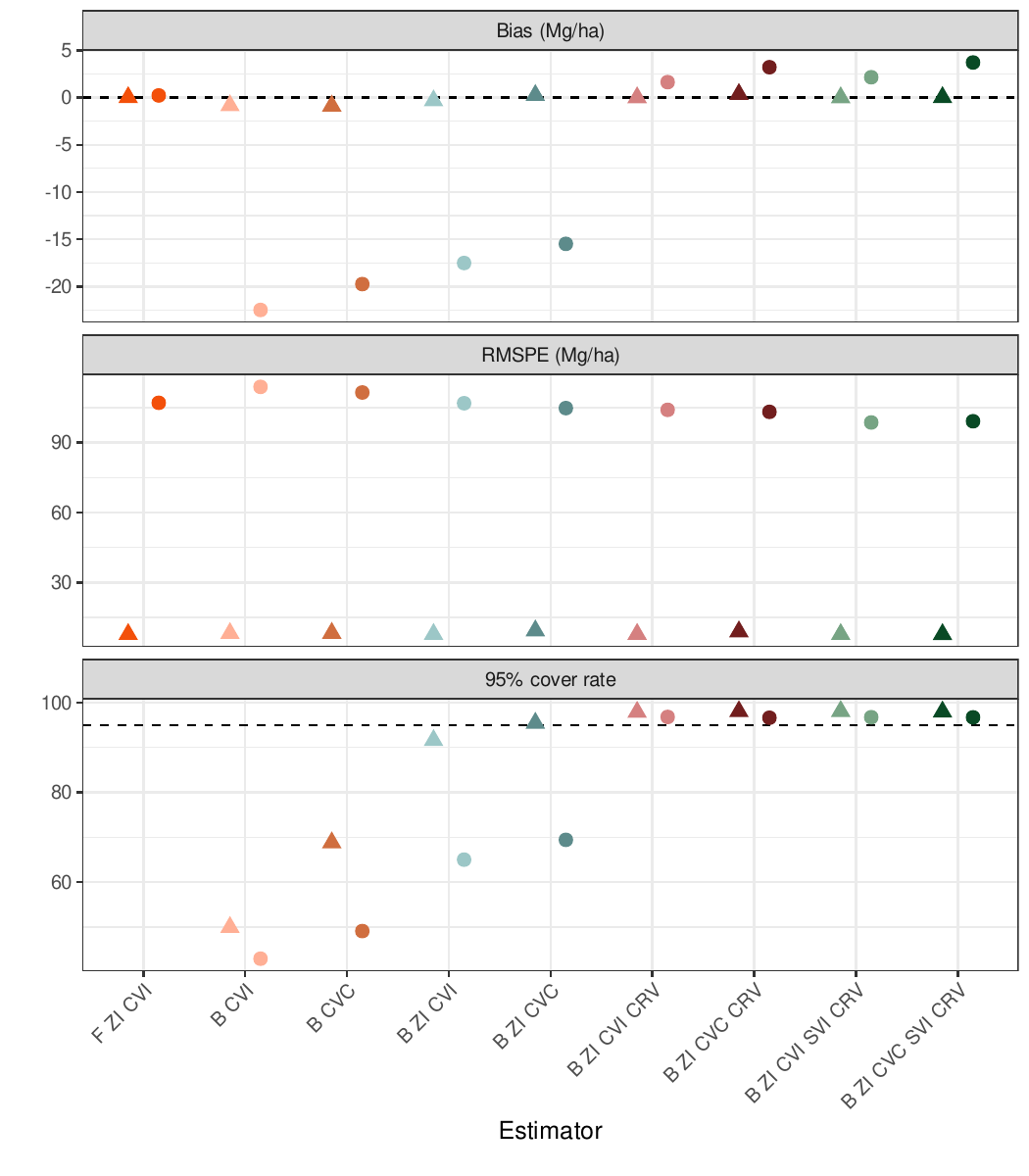}
\caption{Unit-level cross-validation results by state. Panels display empirical bias (Mg/ha), root mean square prediction error (RMSPE; Mg/ha), and 95\% uncertainty interval coverage for each estimator. Nevada (NV) is shown with filled triangles ($\blacktriangle$) and Washington (WA) with filled circles ($\bullet$). Abbreviations: frequentist (F); zero-inflated (ZI); county-varying intercept (CVI); Bayesian (B); county-varying coefficient (CVC); county-specific residual variance (CRV); space-varying intercept (SVI). Dashed horizontal line indicates nominal 95\% coverage.}
 \label{fig:cross_validation}
\end{figure}

Cross-validation was conducted at the unit level, with results summarized in Figure~\ref{fig:cross_validation}. Patterns are broadly consistent with the simulation study. In Nevada, bias was relatively small across all estimators, with two-stage models performing marginally better than single-stage models. In Washington, F ZI CVI showed the smallest bias, followed by the two-stage Bayesian models with county-varying intercepts and residual variances (B ZI CVI CRV and B ZI CVI SVI CRV).

For RMSPE (Figure~\ref{fig:cross_validation}), prediction accuracy was similar across models within each state. In Washington, the Bayesian models with a space-varying intercept term (SVI models) showed marginal improvements in predictive performance.

Coverage rates generally improved with model complexity, with the best performance observed for models that included county-varying residual terms. Coverage is not reported for the frequentist estimator because unit-level prediction intervals are not available in closed form or via bootstrap for the two-stage model.

Simulation results in Section~\ref{sec:simulation-res} and FIA cross-validation results (Figure~\ref{fig:cross_validation}) show that B ZI CVI SVI CRV performed well in both states and across candidate models, so this model was used to generate pixel- and county-level biomass estimates.

Model parameter estimates for this model are provided in Appendix Tables~\ref{tab:nv_parameters} and~\ref{tab:wa_parameters} and Figures~\ref{fig:nv_parameters} and~\ref{fig:wa_parameters} for Nevada and Washington, respectively. Because our focus is on predictive performance, and given the challenges of interpreting regression coefficients in the presence of random effects---particularly spatial random effects \citep[e.g.,][]{Zimmerman2022, Makinen2022, Gilbert2024}---we do not emphasize interpretation of individual coefficients. The county- and space-varying effects shown in Figures~\ref{fig:nv_parameters} and~\ref{fig:wa_parameters} reveal variability and patterns consistent with the spatial heterogeneity of forest and biomass distributions. In particular, the spatial random effects in Washington, and to a lesser degree in Nevada, identify expected patterns of biomass distribution and compensate for missing predictors and the limitations of county-level random effects. While these figures highlight the potential importance of SVI terms, the formal cross-validation metrics show only marginal improvements in RMSPE for the spatial models.

\begin{figure}[ht!]
    \includegraphics[width=\textwidth]{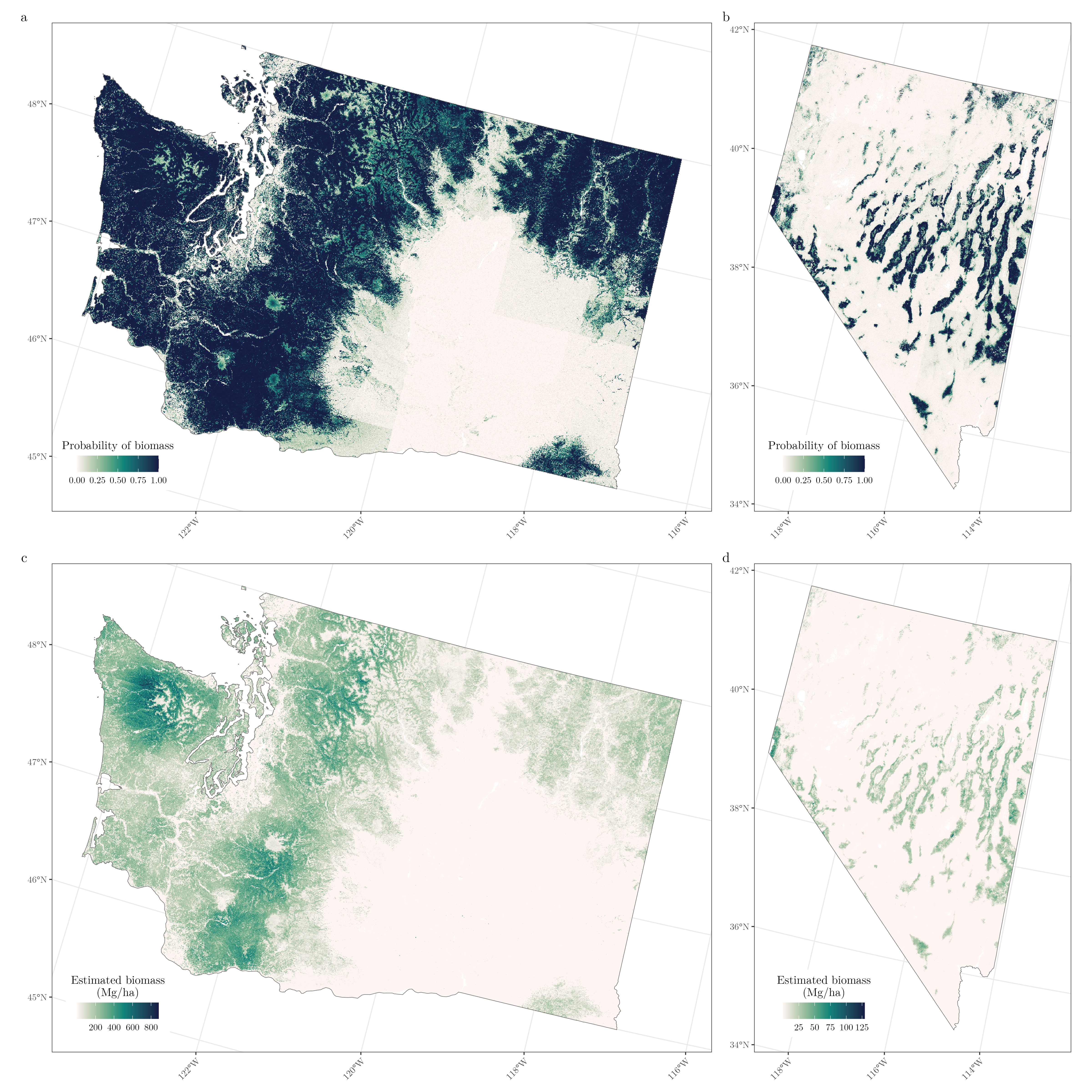}
    \caption{Pixel-level estimates of biomass probability (a, b) and estimated biomass (c, d) in Washington (a, c) and Nevada (b, d). Biomass probabilities are derived from the Bayesian Bernoulli model with county-varying intercepts; biomass estimates are derived from the Bayesian zero-inflated model with county-varying intercept, space-varying intercept, and county-specific residual variances.}
    \label{fig:pixel_quartet}
\end{figure}

Figure~\ref{fig:pixel_quartet} shows pixel-level biomass probabilities (a–b) and biomass estimates (c–d) for both states. These maps illustrate how the modeling framework supports estimation and uncertainty quantification at arbitrary spatial scales, and how each model stage contributes to the results. In Washington, the pixel-level estimates reflect forest distribution and biomass density, with a clear gradient across the Cascade Range. In Nevada, the pixel-level estimates highlight the sky islands and extensive non-forest areas.

\begin{figure}[ht!]
    \includegraphics[width=\textwidth]{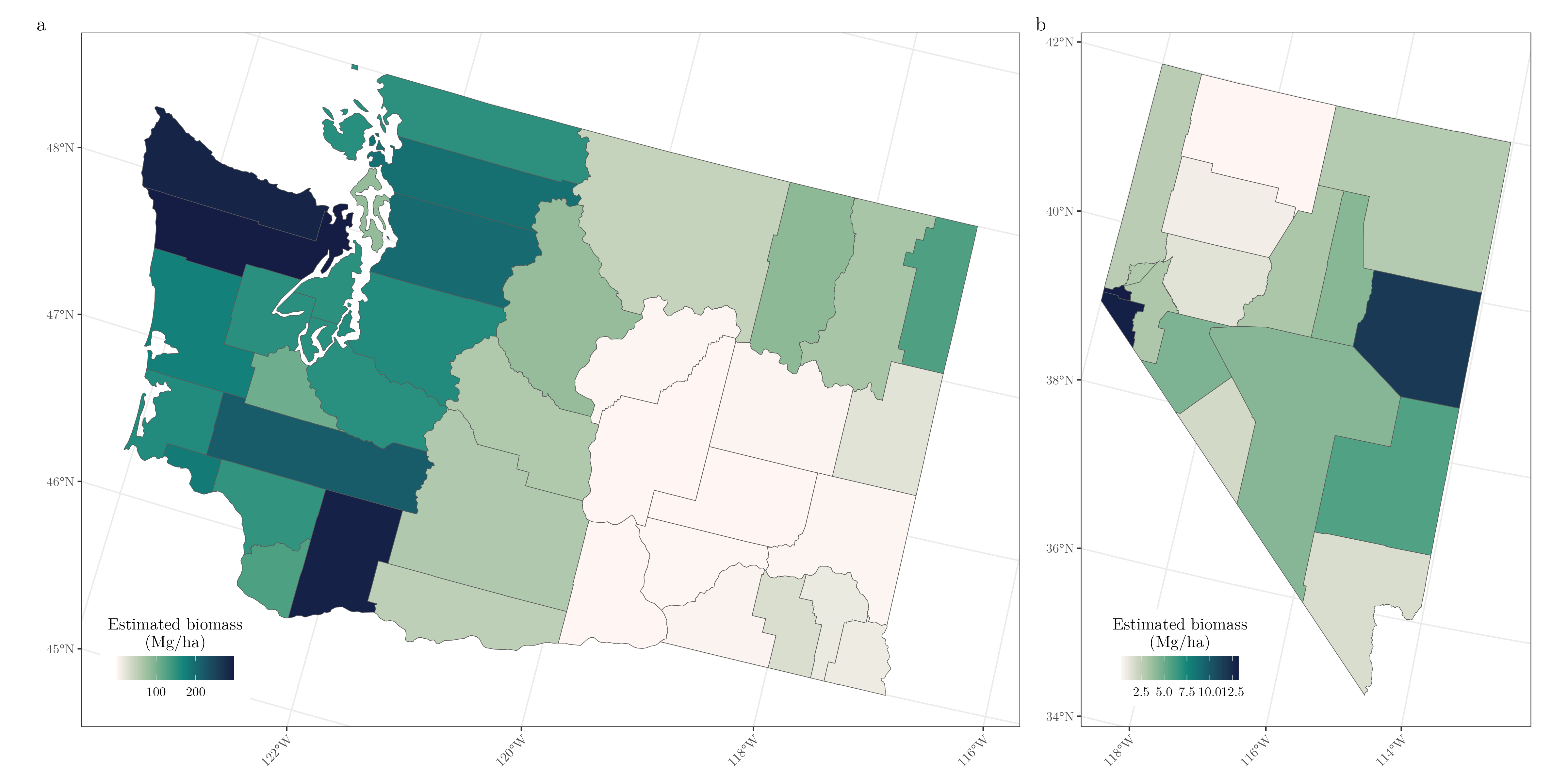}
    \caption{County-level estimates of average biomass (Mg/ha) in Washington (a) and Nevada (b). Estimates are produced from the Bayesian zero-inflated model with county-varying intercept, space-varying intercept, and county-specific residual variances.}
    \label{fig:chloropleth}
\end{figure}

As described in Section~\ref{sec:bayes_pred}, the pixel-level posterior distributions (whose means are shown in Figure~\ref{fig:pixel_quartet}(c–d)) were summarized to produce the county-level $\mu_j$ estimates shown in Figure~\ref{fig:chloropleth}.

\section{Discussion}\label{sec:discussion}

This study evaluated nine model-based estimators for county-level biomass in Nevada and Washington, focusing on how zero-inflation, county-level effects, residual variance structures, and spatial components influence estimator performance. We used two complementary approaches: repeated sampling from simulated populations, which allowed direct comparison to known county-level values, and unit-level cross-validation using FIA data, which assessed predictive performance under real-world conditions. The latter approach, by necessity, evaluated estimator performance at the unit level rather than for the area-level parameters that are the target of inference.

Across both analyses, two-stage models that explicitly accounted for zero-inflation outperformed single-stage models. Incorporating county-specific residual variances (CRVs) consistently reduced bias and RMSE, particularly in Nevada, where biomass is concentrated on isolated sky islands. Adding spatial random effects (SVIs) further improved performance in the FIA analysis, especially in Washington, where biomass patterns are strongly structured by the Cascade Range. In contrast, the benefits of including SVIs were less pronounced in the simulation study, suggesting that the simulated populations may not fully capture the degree of spatial autocorrelation present in the FIA data.

This limitation has been noted previously. \cite{kbaabbrpkg} highlighted that \texttt{kbaabb}-generated populations lack the degree of spatial autocorrelation commonly observed in FIA data. In our FIA analyses, the estimated effective spatial range for the SVI components was approximately 114 km in Washington and 7 km in Nevada (Tables~\ref{tab:wa_parameters} and \ref{tab:nv_parameters}, respectively), confirming the presence of meaningful residual spatial structure in the observed data. Future simulation work that incorporates realistic spatial dependence---for example, through nonparametric smoothing or spatial processes---would enable more realistic assessments of spatial estimators.

Single-stage estimators showed the highest bias and poorest coverage overall. CVC effects tended to increase bias in Nevada without clear benefits, and in Washington they provided only modest improvements in coverage. The frequentist two-stage estimator performed reasonably well but exhibited a pronounced positive bias in RMSE estimation, leading to overly wide confidence intervals and inflated coverage rates. Refining its MSE estimator would improve its utility as a baseline approach. Future work may include refining or developing new MSE estimators for the frequentist two-stage estimator. 

The FIA cross-validation results reinforced the advantages of including CRV and SVI terms. Models combining these components achieved the lowest RMSPE and bias, with coverage rates approaching nominal levels. The frequentist and more complex Bayesian zero-inflated estimators were particularly effective in Nevada, where extensive non-forest areas result in many true zeros that single-stage models cannot accommodate.

Taken together, the simulation and FIA analyses highlight how model structure influences estimator performance and underscore the importance of accounting for key population features when applying SAE methods to forest inventory data.

\section{Conclusion}\label{sec:conclusion}

This study demonstrates that explicitly accommodating zero-inflation, variance heterogeneity at the county level, and residual spatial structure can be important for producing accurate and well-calibrated biomass estimates across contrasting ecological settings. Bayesian two-stage models provide a flexible framework for integrating these components, offering improved calibration, reduced bias, and robust uncertainty propagation compared with simpler alternatives.

The simulation and FIA cross-validation analyses provided complementary perspectives. While similar trends in estimator performance were seen between the simulated county-level and FIA unit-level assessments, the county-level assessments are better suited to model selection because the target parameters are defined at the county level. Together, these approaches offer a more complete view of estimator behavior under both controlled and real-world conditions.

The results offer practical guidance for producing reliable county-level biomass estimates, which are increasingly needed for resource monitoring, wildfire risk assessment, and local-scale planning. By improving population generation methods to better reflect spatial structure and extending these models to additional spatial and temporal domains, future work can further strengthen the role of model-based SAE methods in forest inventory and environmental monitoring.

\clearpage
\bibliographystyle{apalike}
\bibliography{references}

@book{congress2022,
title={S. Rep. 118-83 - Department of the Interior, Environment, and Related Agencies Appropriations Bill}, 
year={2023},
author={{U.S. Senate}},
url = {https://www.govinfo.gov/app/details/CRPT-118srpt83/context}}

@article{pfeffermann,
author = {Pfeffermann, Danny and Terryn, Bénédicte and  Moura, Fernando},
year = {2008},
month = {12},
pages = {},
title = {Small area estimation under a two-part random effects model with application to estimation of literacy in developing countries},
volume = {34},
journal = {Survey Methodology}
}

@article{chandra_sud,
  title={Small Area Estimation for Zero-Inflated Data},
  author={Hukum Chandra and U. C. Sud},
  journal={Communications in Statistics - Simulation and Computation},
  year={2012},
  volume={41},
  pages={632 - 643}
}

@article{finley2011hierarchical,
  title={A hierarchical model for quantifying forest variables over large heterogeneous landscapes with uncertain forest areas},
  author={Finley, Andrew O and Banerjee, Sudipto and MacFarlane, David W},
  journal={Journal of the American Statistical Association},
  volume={106},
  number={493},
  pages={31--48},
  year={2011},
  publisher={Taylor \& Francis}
}

@article{white2024small,
author = {White, Grayson W. and Yamamoto, Josh K. and Elsyad, Dinan H. and Schmitt, Julian F. and Korsgaard, Niels H. and Hu, Jie Kate and Gaines, George C. and Frescino, Tracey S. and McConville, Kelly S.},
title = {Small area estimation of forest biomass via a two-stage model for continuous zero-inflated data},
journal = {Canadian Journal of Forest Research},
volume = {55},
number = {},
pages = {1-19},
year = {2025},
doi = {10.1139/cjfr-2024-0149},

URL = { 
    
        https://doi.org/10.1139/cjfr-2024-0149
    
    

},
eprint = { 
    
        https://doi.org/10.1139/cjfr-2024-0149
    
    

}
}

@article{fay1979estimates,
  title={Estimates of income for small places: an application of James-Stein procedures to census data},
  author={Fay, Robert E III and Herriot, Roger A},
  journal={Journal of the American Statistical Association},
  volume={74},
  number={366a},
  pages={269--277},
  year={1979},
  publisher={Taylor \& Francis}
}

@article{battese1988error,
  title={An error-components model for prediction of county crop areas using survey and satellite data},
  author={Battese, George E and Harter, Rachel M and Fuller, Wayne A},
  journal={Journal of the American Statistical Association},
  volume={83},
  number={401},
  pages={28--36},
  year={1988},
  publisher={Taylor \& Francis}
}

@Book{rao15,
    title = {Small Area Estimation},
    author = {Rao, J.N.K. and Molina, Isabel},
    year = {2015},
    publisher = {Wiley},
    edition = {2nd},
    note = {ISBN: 978-1-118-73578-7}
}

@misc{white2024assessing,
      title={A method for empirically assessing small area estimators via bootstrap-weighted k-Nearest-Neighbor artificial populations, with applications to forest inventory}, 
      author={Grayson W. White and Jerzy A. Wieczorek and Zachariah W. Cody and Emily X. Tan and Jacqueline O. Chistolini and Kelly S. McConville and Tracey S. Frescino and Gretchen G. Moisen},
      year={2025},
      eprint={2306.15607},
      archivePrefix={arXiv},
      primaryClass={stat.ME}
}

@article{yang2018new,
  title={A new generation of the {U}nited {S}tates {N}ational {L}and {C}over {D}atabase: Requirements, research priorities, design, and implementation strategies},
  author={Yang, Limin and Jin, Suming and Danielson, Patrick and Homer, Collin and Gass, Leila and Bender, Stacie M and Case, Adam and Costello, Catherine and Dewitz, Jon and Fry, Joyce and Funk, Michelle and Granneman, Brian and Liknes, Greg C and Rigge, Matthew and Xian, George},
  journal={ISPRS Journal of Photogrammetry and Remote Sensing},
  volume={146},
  pages={108--123},
  year={2018},
  publisher={Elsevier}
}

@misc{usgs2019ned,
    author={{U.S. Geological Survey}}, 
    title={{LANDFIRE} {E}levation}, 
    location={USGS EROS, Sioux Falls, South Dakota}, 
    year={2019}
}

@article{daly2002knowledge,
  title={A knowledge-based approach to the statistical mapping of climate},
  author={Daly, Christopher and Gibson, Wayne P and Taylor, George H and Johnson, Gregory L and Pasteris, Phillip},
  journal={Climate Research},
  volume={22},
  number={2},
  pages={99--113},
  year={2002}
}

@article{picotte2019landfire,
  title={{LANDFIRE} remap prototype mapping effort: Developing a new framework for mapping vegetation classification, change, and structure},
  author={Picotte, Joshua J and Dockter, Daryn and Long, Jordan and Tolk, Brian and Davidson, Anne and Peterson, Birgit},
  journal={Fire},
  volume={2},
  number={2},
  pages={35},
  year={2019},
  publisher={MDPI}
}

@article{rollins2009landfire,
  title={{LANDFIRE}: a nationally consistent vegetation, wildland fire, and fuel assessment},
  author={Rollins, Matthew G},
  journal={International Journal of Wildland Fire},
  volume={18},
  number={3},
  pages={235--249},
  year={2009},
  publisher={CSIRO Publishing}
}

@Manual{kbaabbrpkg,
    title = {kbaabb: Generates an Artificial Population Based on the KBAABB
Methodology},
    author = {Grayson W. White and Jerzy A. Wieczorek and Tracey S. Frescino and Kelly S. McConville},
    year = {2024},
    note = {R package version 0.0.0.9000},
  }

@article{wiener2021united,
  title={United States forest service use of forest inventory data: Examples and needs for small area estimation},
  author={Wiener, Sarah S and Bush, Renate and Nathanson, Amy and Pelz, Kristen and Palmer, Marin and Alexander, Mara L and Anderson, David and Treasure, Emrys and Baggs, Joanne and Sheffield, Ray},
  journal={Frontiers in Forests and Global Change},
  volume={4},
  pages={763487},
  year={2021},
  publisher={Frontiers Media SA}
}

@article{prisley2021needs,
  title={Needs for small area estimation: Perspectives from the US private forest sector},
  author={Prisley, Steve and Bradley, Jeff and Clutter, Mike and Friedman, Suzy and Kempka, Dick and Rakestraw, Jim and Sonne Hall, Edie},
  journal={Frontiers in Forests and Global Change},
  volume={4},
  pages={746439},
  year={2021},
  publisher={Frontiers Media SA}
}

@article{shannon2024toward,
  title={Toward spatio-temporal models to support national-scale forest carbon monitoring and reporting},
  author={Shannon, Elliot S and Finley, Andrew O and Domke, Grant M and May, Paul B and Andersen, Hans-Erik and Gaines III, George C and Banerjee, Sudipto},
  journal={Environmental Research Letters},
  volume={20},
  number={1},
  pages={014052},
  year={2024},
  publisher={IOP Publishing}
}

@article{may2023spatially,
  title={A spatially varying model for small area estimates of biomass density across the contiguous United States},
  author={May, Paul and McConville, Kelly S and Moisen, Gretchen G and Bruening, Jamis and Dubayah, Ralph},
  journal={Remote Sensing of Environment},
  volume={286},
  pages={113420},
  year={2023},
  publisher={Elsevier}
}

@article{cao2022increased,
  title={Increased precision in county-level volume estimates in the United States National Forest Inventory with area-level small area estimation},
  author={Cao, Qianqian and Dettmann, Garret T and Radtke, Philip J and Coulston, John W and Derwin, Jill and Thomas, Valerie A and Burkhart, Harold E and Wynne, Randolph H},
  journal={Frontiers in Forests and Global Change},
  volume={5},
  pages={769917},
  year={2022},
  publisher={Frontiers Media SA}
}

@article{finley2024models,
  title={Models to support forest inventory and small area estimation using sparsely sampled LiDAR: A case study involving G-LiHT LiDAR in Tanana, Alaska},
  author={Finley, Andrew O and Andersen, Hans-Erik and Babcock, Chad and Cook, Bruce D and Morton, Douglas C and Banerjee, Sudipto},
  journal={Journal of Agricultural, Biological and Environmental Statistics},
  volume={29},
  number={4},
  pages={695--722},
  year={2024},
  publisher={Springer}
}

@misc{burrillforest,
    title={Forest Inventory and Analysis Database: Database Description and User Guide for Phase 2 (version: 9.1)},
    author={Burrill, E.A. and Christensen, G.A. and Conkling, B.L. and DiTommaso, A.M. and Lepine, L. and Perry, C.J. and Pugh, S.A. and Turner, J.A. and Walker, D. and Williams, M.A.},
    year={2023},
    url={https://www.fs.usda.gov/research/understory/forest-inventory-and-analysis-database-user-guide-nfi}
}

@article{datta2016hierarchical,
  title={{Hierarchical nearest-neighbor Gaussian process models for large geostatistical datasets}},
  author={Datta, Abhirup and Banerjee, Sudipto and Finley, Andrew O and Gelfand, Alan E},
  journal={Journal of the American Statistical Association},
  volume={111},
  number={514},
  pages={800--812},
  year={2016},
  publisher={Taylor \& Francis}
}

@article{finley2019efficient,
  title={{Efficient algorithms for Bayesian nearest neighbor Gaussian processes}},
  author={Finley, Andrew O and Datta, Abhirup and Cook, Bruce D and Morton, Douglas C and Andersen, Hans E and Banerjee, Sudipto},
  journal={Journal of Computational and Graphical Statistics},
  volume={28},
  number={2},
  pages={401--414},
  year={2019},
  publisher={Taylor \& Francis}
}

@Manual{saeczi,
    title = {saeczi: Small Area Estimation for Continuous Zero Inflated Data},
    author = {Josh Yamamoto and Dinan Elsyad and Grayson White and Julian Schmitt and Niels Korsgaard and Kelly McConville and Kate Hu},
    year = {2025},
    note = {R package version 0.2.0.9000, commit f580319049b152c890033c770913b7a296ce63cb},
    url = {https://github.com/joshyam-k/saeczi-dev},
  }

@Manual{R,
    title = {R: A Language and Environment for Statistical Computing},
    author = {{R Core Team}},
    organization = {R Foundation for Statistical Computing},
    address = {Vienna, Austria},
    year = {2024},
    url = {https://www.R-project.org/},
  }

@Manual{ZICode,
    title = {Models for zero-inflated data},
    author = {Andrew O. Finley},
    year = {2025},
    note = {Available at \url{https://github.com/finleya/zi_models}},
  }

@book{gelman2013bayesian,
  title={Bayesian Data Analysis, Third Edition},
  author={Gelman, A. and Carlin, J.B. and Stern, H.S. and Dunson, D.B. and Vehtari, A. and Rubin, D.B.},
  isbn={9781439840955},
  lccn={2013039507},
  series={Chapman \& Hall/CRC Texts in Statistical Science},
  year={2013},
  publisher={Taylor \& Francis}
}

@article{breidenbach2012small,
  title={Small area estimation of forest attributes in the Norwegian National Forest Inventory},
  author={Breidenbach, Johannes and Astrup, Rasmus},
  journal={European Journal of Forest Research},
  volume={131},
  number={4},
  pages={1255--1267},
  year={2012},
  publisher={Springer}
}

@article{kangas,
author = {Kangas, Annika and Myllymäki, Mari and Packalen, Petteri},
title = {Small area estimators in a simulation test},
journal = {Canadian Journal of Forest Research},
volume = {55},
number = {},
pages = {1-17},
year = {2025},
doi = {10.1139/cjfr-2024-0070},
URL = {https://doi.org/10.1139/cjfr-2024-0070},
eprint = {https://doi.org/10.1139/cjfr-2024-0070}
}

@article{Nothdurft2025,
author = {Nothdurft, Arne and Sarkleti, Valentin and Ofner-Graff, Tobias and Tockner, Andreas and Gollob, Christoph and Ritter, Tim and Kraßnitzer, Ralf and Svazek, Philip and Kühmaier, Martin and Stampfer, Karl and Finley, Andrew O.},
title = {Small area estimation of growing stock timber volume, basal area, mean stem diameter, and stem density for mountain forests in Austria},
journal = {Canadian Journal of Forest Research},
volume = {55},
number = {},
pages = {1-20},
year = {2025},
doi = {10.1139/cjfr-2024-0302}
}

@article{SHANNON2025122999,
title = {Leveraging national forest inventory data to estimate forest carbon density status and trends for small areas},
journal = {Forest Ecology and Management},
volume = {596},
pages = {122999},
year = {2025},
issn = {0378-1127},
doi = {https://doi.org/10.1016/j.foreco.2025.122999},
url = {https://www.sciencedirect.com/science/article/pii/S0378112725005079},
author = {Elliot S. Shannon and Andrew O. Finley and Paul B. May and Grant M. Domke and Hans-Erik Andersen and George C. {Gaines III} and Arne Nothdurft and Sudipto Banerjee}
}

@article{gregoire2008regression,
  title={Regression estimation following the square-root transformation of the response},
  author={Gregoire, Timothy G and Lin, Qi Feng and Boudreau, Johnathan and Nelson, Ross},
  journal={Forest Science},
  volume={54},
  number={6},
  pages={597--606},
  year={2008},
  publisher={Oxford University Press}
}

@article{stow2006bayesian,
  title={A Bayesian approach to retransformation bias in transformed regression},
  author={Stow, Craig A and Reckhow, Kenneth H and Qian, Song S},
  journal={Ecology},
  volume={87},
  number={6},
  pages={1472--1477},
  year={2006},
  publisher={Wiley Online Library}
}

@article{bates2015package,
  title={Package ‘lme4’},
  author={Bates, Douglas and Maechler, Martin and Bolker, Ben and Walker, Steven and Christensen, Rune Haubo Bojesen and Singmann, Henrik and Dai, Bin and Grothendieck, Gabor and Green, Peter and Bolker, Maintainer Ben},
  journal={convergence},
  volume={12},
  number={1},
  pages={2},
  year={2015}
}

@article{Makinen2022,
author = {Mäkinen, Jussi and Numminen, Elina and Niittynen, Pekka and Luoto, Miska and Vanhatalo, Jarno},
title = {Spatial confounding in Bayesian species distribution modeling},
journal = {Ecography},
volume = {2022},
number = {11},
pages = {e06183},
doi = {https://doi.org/10.1111/ecog.06183},
url = {https://nsojournals.onlinelibrary.wiley.com/doi/abs/10.1111/ecog.06183},
eprint = {https://nsojournals.onlinelibrary.wiley.com/doi/pdf/10.1111/ecog.06183},
year = {2022}
}

@article{Zimmerman2022,
author = {Dale L. Zimmerman and Jay M. Ver Hoef},
title = {On Deconfounding Spatial Confounding in Linear Models},
journal = {The American Statistician},
volume = {76},
number = {2},
pages = {159--167},
year = {2022},
publisher = {ASA Website},
doi = {10.1080/00031305.2021.1946149}
}

@article{Gilbert2024,
    author = {Gilbert, Brian and Ogburn, Elizabeth L and Datta, Abhirup},
    title = {Consistency of common spatial estimators under spatial confounding},
    journal = {Biometrika},
    volume = {112},
    number = {2},
    pages = {asae070},
    year = {2024},
    month = {12}
}

@article{tzavidis2018start,
  title={From start to finish: a framework for the production of small area official statistics},
  author={Tzavidis, Nikos and Zhang, Li-Chun and Luna, Angela and Schmid, Timo and Rojas-Perilla, Natalia},
  journal={Journal of the Royal Statistical Society, Series A (Statistics in Society)},
  volume={181},
  number={4},
  pages={927--979},
  year={2018}
}

\clearpage

\appendix
\renewcommand{\thefigure}{A.\arabic{figure}}  
\renewcommand{\thetable}{A.\arabic{table}}    
\setcounter{figure}{0}                        
\setcounter{table}{0}                         

\section{Appendix}\label{sec:appendixA}


Table~\ref{tab:prior_distributions} includes the prior distribution and hyperparameter values for all parameters used for the Bayesian models. 

\begin{table}[ht!]
\centering
\begin{tabular}{c|c|c}
\hline
Parameter & Prior distribution & Hyperparameter values \\
\hline
$\alpha_0$ & $\mathcal{N}(\mu,~\sigma^2)$ & $\mu = 0,~ \sigma^2 = 1000$ \\
$\tilde\alpha_{0,j}$ & $\mathcal{N}(\mu,~\sigma^2)$ & $\mu = 0,~ \sigma^2 = \sigma^2_{\tilde\alpha_0}$ \\ 
$\alpha_k$ & $\mathcal{N}(\mu,~\sigma^2)$ & $\mu = 0,~\sigma^2 = 1000$ \\ 
$\sigma^2_{\tilde\alpha_0}$ & $\mathcal{IG}(a,~b)$ & $a = 2,~ b = 1$ \\
$\beta_0$ & $\mathcal{N}(\mu,~\sigma^2)$ & $\mu = 0,~ \sigma^2 = 1000$ \\
$\tilde\beta_{0,j}$ & $\mathcal{N}(\mu,~\sigma^2)$ & $\mu = 0,~ \sigma^2 = \sigma^2_{\tilde\beta_0}$ \\ 
$\beta_k$ & $\mathcal{N}(\mu,~\sigma^2)$ & $\mu = 0,~\sigma^2 = 1000$ \\ 
$\tilde\beta_{k,j}$ & $\mathcal{N}(\mu,~\sigma^2)$ & $\mu =0,~ \sigma^2 = \sigma^2_{\tilde\beta_{k,j}}$ \\
$\sigma^2_{\tilde\beta_0}$ & $\mathcal{IG}(a,~b)$ & $a = 2,~ b = 1$ \\
$\sigma^2_{\tilde\beta_{k,j}}$ & $\mathcal{IG}(a,~b)$ & $a = 2,~ b = 1$ \\

$\sigma^2_w$ & $\mathcal{IG}(a,~b)$ & $a = 2,~ b = 1$ \\
$\phi$ & $\mathcal{U}(a,~b)$ & $a = 0.003,~ b = 3$ \\
$\tau^2$ & $\mathcal{IG}(a,~b)$ &$a=2,~b=1$ \\
$\tau_1^2$ & $\mathcal{IG}(a,~b)$ &$a=2,~b=1$ \\
$\tau_{1,j}^2$ & $\mathcal{IG}(a,~b)$ &$a=2,~b=1$ \\
\hline
\end{tabular}
\caption{Prior distributions and hyperparameter values used for the Bayesian models.} \label{tab:prior_distributions} 
\end{table}

\begin{table}[ht]
\centering
\begin{tabular}{l c}
\hline
\multicolumn{2}{c}{Bernoulli model for $z(\ell)$} \\
\hline
Parameter & Posterior mean (95\% CrI) \\
\hline
$\alpha_0$ & -6.410 $(-7.023,\ -5.794)$ \\
$\alpha_{\text{tcc}}$ & 0.148 $(0.114,\ 0.180)$ \\
$\alpha_{\text{elev}}$ & -0.010 $(-0.019,\ 0.000)$ \\
$\alpha_{\text{tri}}$ & 0.268 $(0.253,\ 0.284)$ \\
$\sigma^2_{\tilde{\alpha}_0}$ & 0.389 $(0.191,\ 0.756)$ \\
\hline
\multicolumn{2}{c}{Continuous model for $y(\ell)$} \\
\hline
Parameter & Posterior mean (95\% CrI) \\
\hline
$\beta_0$ & 1.450 $(0.755,\ 2.188)$ \\
$\beta_{\text{tcc}}$ & 0.058 $(0.024,\ 0.094)$ \\
$\beta_{\text{elev}}$ & -0.004 $(-0.012,\ 0.003)$ \\
$\beta_{\text{tri}}$ & 0.080 $(0.074,\ 0.086)$ \\
$\sigma^2_{\tilde{\beta}_0}$ & 0.259 $(0.119,\ 0.519)$ \\
$\sigma^2_{w}$ & 1.502 $(1.218,\ 1.742)$ \\
Eff. rng. (km) & 6.679 $(5.189,\ 8.434)$ \\
\hline
\end{tabular}
\caption{Posterior means and 95\% credible intervals (CrI) for parameters of the B ZI CVI SVI CRV model \eqref{mod:bayes_zi_CVI_SVI_CRV} fit to FIA data in Nevada. The Bernoulli model corresponds to biomass presence $z(\ell)$, and the continuous model corresponds to transformed biomass $y(\ell)$. Estimates for the remaining model parameters are shown in Figure~\ref{fig:nv_parameters} because they vary over counties or across a continuous spatial surface and are more effectively visualized as maps. The effective range (Eff. rng.) is computed using estimates for $\phi$ as $-\log(0.05)/\phi$ and represents the distance (km) at which spatial correlation drops to 0.05.}
\label{tab:nv_parameters}
\end{table}

\begin{table}[ht]
\centering
\begin{tabular}{l c}
\hline
\multicolumn{2}{c}{Bernoulli model for $z(\ell)$} \\
\hline
Parameter & Posterior mean (95\% CrI) \\
\hline
$\alpha_0$ & -4.785 $(-5.258,\ -4.387)$ \\
$\alpha_{\text{tcc}}$ & 0.124 $(0.097,\ 0.151)$ \\
$\alpha_{\text{elev}}$ & 0.014 $(-0.006,\ 0.034)$ \\
$\alpha_{\text{ppt}}$ & 0.102 $(0.095,\ 0.109)$ \\
$\alpha_{\text{tnt}}$ & 1.222 $(0.989,\ 1.460)$ \\
$\sigma^2_{\tilde{\alpha}_0}$ & 0.863 $(0.487,\ 1.424)$ \\
\hline
\multicolumn{2}{c}{Continuous model for $y(\ell)$} \\
\hline
Parameter & Posterior mean (95\% CrI) \\
\hline
$\beta_0$ & 1.752 $(1.606,\ 1.893)$ \\
$\beta_{\text{tcc}}$ & -0.008 $(-0.016,\ 0.001)$ \\
$\beta_{\text{elev}}$ & 0.014 $(0.008,\ 0.019)$ \\
$\beta_{\text{ppt}}$ & 0.024 $(0.022,\ 0.025)$ \\
$\sigma^2_{\tilde{\beta}_0}$ & 0.084 $(0.052,\ 0.134)$ \\
$\sigma^2_{w}$ & 0.157 $(0.112,\ 0.195)$ \\
Eff. rng. (km) & 114.009 $(93.033,\ 137.224)$ \\
\hline
\end{tabular}
\caption{Posterior means and 95\% credible intervals (CrI) for parameters of the B ZI CVI SVI CRV model fit to FIA data in Washington. The Bernoulli model corresponds to biomass presence $z(\ell)$, and the continuous model corresponds to transformed biomass $y(\ell)$. Estimates for the remaining model parameters are shown in Figure~\ref{fig:wa_parameters} because they vary over counties or across a continuous spatial surface and are more effectively visualized as maps. The effective range (Eff. rng.) is computed using estimates for $\phi$ as $-\log(0.05)/\phi$ and represents the distance (km) at which spatial correlation drops to 0.05.}
\label{tab:wa_parameters}
\end{table}

\begin{figure}[ht!]
\centering
    \includegraphics[width=0.5\textwidth]{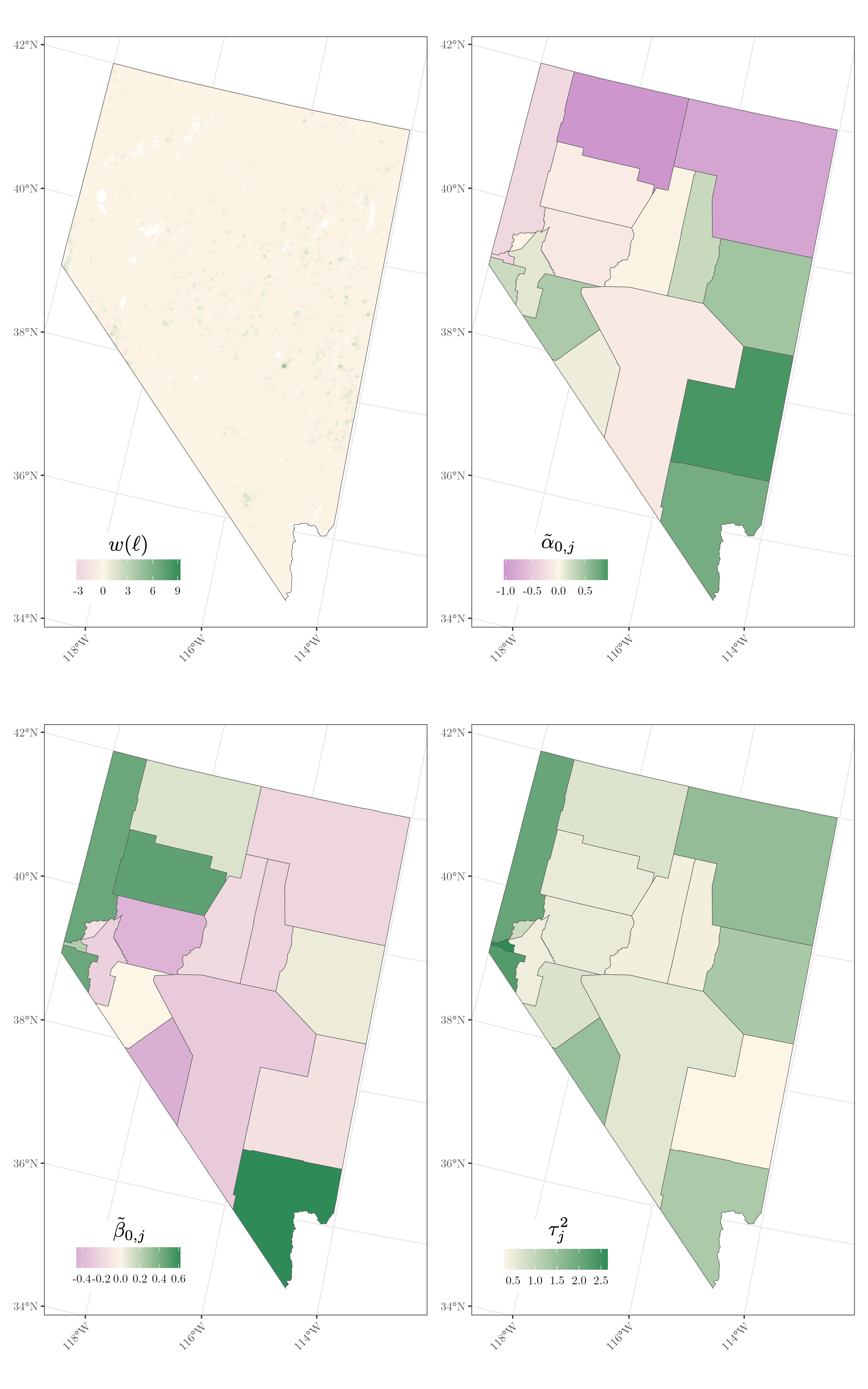}
    \caption{Posterior means for county-level and space-varying parameters of the B ZI CVI SVI CRV model fit to FIA data in Nevada. Each subfigure corresponds to a single parameter, with the posterior mean displayed above the scale bar. Parameter definitions and fixed-effect estimates are provided in Table~\ref{tab:nv_parameters}.}
    \label{fig:nv_parameters}
\end{figure}

\begin{figure}[ht!]
\centering
    \includegraphics[width=\textwidth]{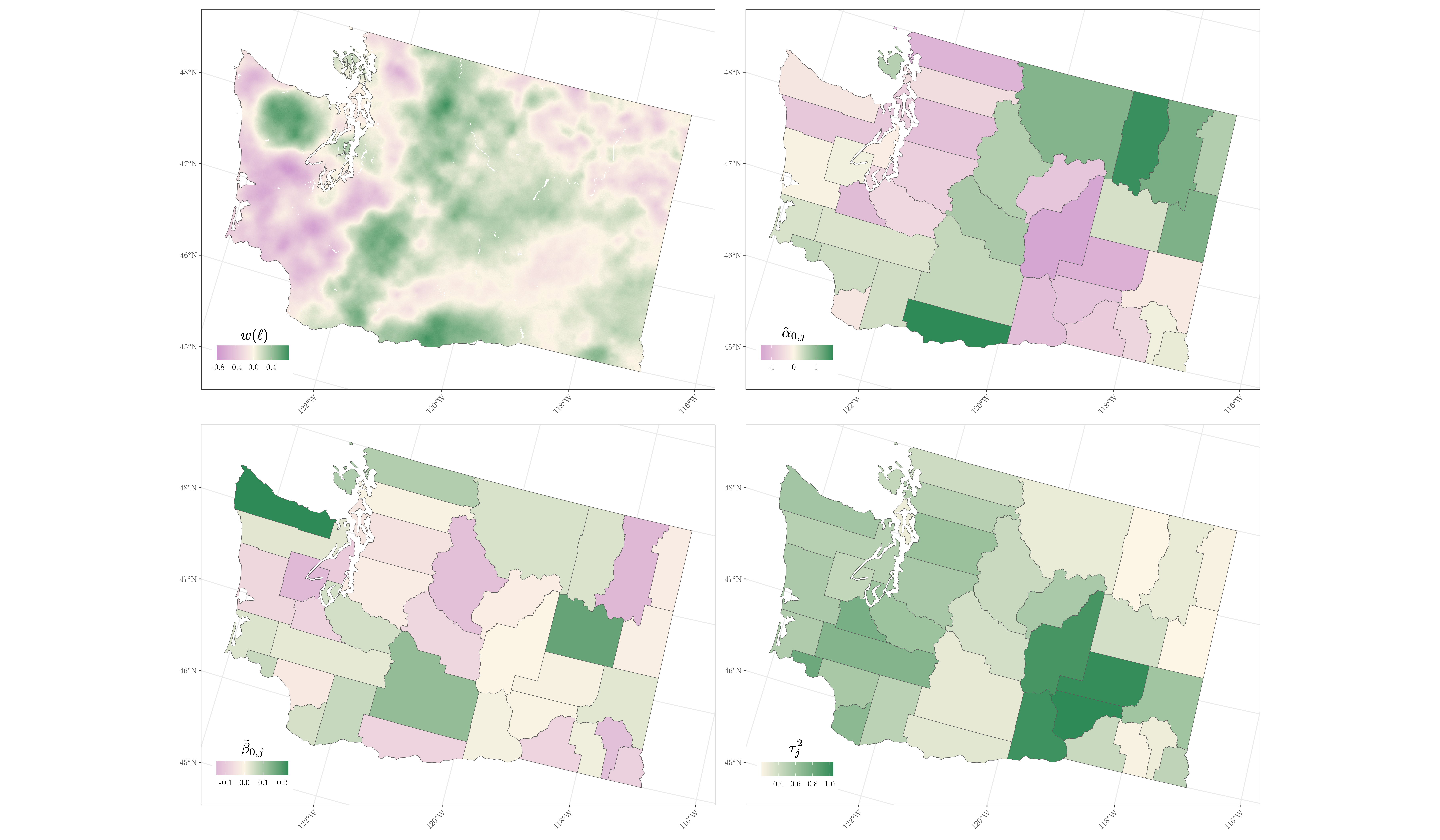}
    \caption{Posterior means for county-level and space-varying parameters of the B ZI CVI SVI CRV model fit to FIA data in Washington. Each subfigure corresponds to a single parameter, with the posterior mean displayed above the scale bar. Parameter definitions and fixed-effect estimates are provided in Table~\ref{tab:wa_parameters}.}
    \label{fig:wa_parameters}
\end{figure}

\end{document}